\documentclass[]{aa}  
\usepackage{graphicx}
\usepackage{color}
\usepackage{epsfig}
\usepackage{supertabular}
\usepackage{aalongtable}
\usepackage{txfonts}
\begin{document}
   \title{The early-type dwarf galaxy population of the Fornax cluster}


   \author{S. Mieske
          \inst{1}
          \and
          M. Hilker\inst{2}\and L. Infante\inst{3} \and C. Mendes de Oliveira\inst{4}
          }

   \offprints{S. Mieske}
   \institute{European Southern Observatory, Karl-Schwarzschild-Strasse 2, 85748 Garching bei M\"unchen, Germany\\
     \email{smieske@eso.org} \and
     Argelander Institut f\"ur Astronomie, Abteilung Sternwarte, Auf dem H\"ugel 71, 53121 Bonn, Germany\\
     \email{mhilker@astro.uni-bonn.de} \and Departamento de
     Astronom\'{\i}a y Astrof\'{\i}sica, Pontificia
     Universidad Cat\'olica de Chile, Casilla 306, Santiago 22, Chile\\
     \email{linfante@astro.puc.cl} \and
     Instituto de Astronomia, Geof\'isica, e Ci\^encias Atmosf\'ericas, Departamento de Astronomia, Universidade de S\~ao Paulo, Rua do Mat\'ao 1226, Cidade Universit\~aria, 05508-900 S\~ao Paulo, SP, Brazil\\
     \email{oliveira@astro.iag.usp.br} }

   \date{}

 
  \abstract
   {} 
   {We analyse the photometric properties of the early-type Fornax
     cluster dwarf galaxy population ($M_{\rm V}>-17$ mag), based on a wide
     field imaging study of the central cluster area in $V$ and $I$
     bandpasses.  We used the instrument/telescope combination
     IMACS/Magellan at Las Campanas Observatory, providing much larger
     light collecting area and better image resolution than previous
     wide field imaging surveys. }
   {We create a fiducial sample of Fornax cluster dwarf ellipticals
     (dEs) in the following three steps: (1) To verify cluster
     membership, we measured $I$-band surface brightness fluctuations
     (SBF) distances to candidate dEs known from previous surveys; (2)
     We re-assessed morphological classifications for those candidate
     dEs that are too faint for SBF detection; and (3) We searched for
     new candidate dEs in the size-luminosity regime close to the
     resolution limit of previous surveys.}
   {(1) We confirm cluster membership for 28 candidate dEs in the
     range $-16.6<M_{\rm V}<-10.1$ mag by means of SBF measurement. We find
     no SBF background galaxy. (2) Of 51 further candidate dEs in the
     range $-13.2<M_{\rm V}<-8.6$ mag, 2/3 are confirmed as probable cluster
     members by morphological re-assessment, while 1/3 are
     re-classified as probable background objects. (3) We find 12 new
     dE candidates in the range $-12.3<M_{\rm V}<-8.8$ mag, two of which are
     directly confirmed via SBF measurement. The resulting fiducial dE
     sample follows a well-defined surface brightness - magnitude
     relation, showing that Fornax dEs are about 40\% larger than
     Local Group dEs. The sample also defines a colour-magnitude
     relation that appears slightly shallower than that of Local Group
     dEs. The early-type dwarf galaxy luminosity function in Fornax
     has a very flat faint end slope $\alpha \simeq -1.1 \pm 0.1$. We
     discuss these findings in the context of structure formation
     theories.}
   {The SBF method is a very powerful tool to help constrain the faint
     end of the galaxy luminosity function in nearby galaxy clusters.
     For the Fornax cluster, morphological cluster memberships -- if
     performed at sufficient resolution -- are very reliable.}
   

   \maketitle
%

\section{Introduction}
\subsection{Fornax cluster galaxy luminosity function}
The Fornax cluster is the most prominent nearby galaxy cluster in the
southern hemisphere, being located at about 19 Mpc ($(m-M)=31.39$ mag)
distance (Freedman et al.~\cite{Freedm01}). Due to its compact nature
and dominance of early-type galaxies (Ferguson~\cite{Fergus89}), the
Fornax cluster is well suited for characterising the global properties
of its galaxy population. The reference source of information on the
Fornax galaxy population is the Fornax Cluster Catalog (FCC,
Ferguson~\cite{Fergus89}), a wide-field imaging survey with 1.68$''$
pixel scale on photographic scans that covers the inner 3.5$\degr$ of
the cluster.

One of the most important quantities for characterising a galaxy
population is the galaxy luminosity function (GLF). Its logarithmic
faint end slope $\alpha$ is a very useful quantity to be contrasted
with the expected slope for the mass spectrum of cosmological dark
matter halos (e.g. Jenkins et al.~\cite{Jenkin01}, Moore et
al.~\cite{Moore99}). Generally, the value of $\alpha$ derived in
various environments including the Local Group is much shallower than
the expected slope of dark matter halos (see for example Grebel et
al.~\cite{Grebel03}; Trentham \& Tully~\cite{Trenth02}; Trentham et
al.~\cite{Trenth05}; Andreon et al.~\cite{Andreo06}; Tanaka et
al.~\cite{Tanaka04}; and Infante et al.~\cite{Infant02} and references
therein). This discrepancy is also known as the ``substructure
problem'' of present day cosmology.

Up to now, investigations of the Fornax GLF in the low luminosity
regime ($M_{\rm V}>-14$ mag) have been restricted to morphological cluster
membership assignment (e.g. Caldwell~\cite{Caldwe87}, Ferguson \&
Sandage~\cite{Fergus88}, Ferguson~\cite{Fergus89}, Phillipps et
al.~\cite{Philli87}, Kambas et al.~\cite{Kambas00}, Hilker et
al.~\cite{Hilker03}). This is because spectroscopic surveys have not
had the sufficient depth to obtain velocities for dE candidates
fainter than about $\mu_{\rm V,0} \simeq$ 23 mag /arcsec$^2$ (e.g. Hilker
et al.~\cite{Hilker99}, Drinkwater et al.~\cite{Drinkw01}). The faint
end slopes derived for the Fornax GLF in the literature generally
cover the range $-1.5<\alpha<-1.0$, depending on the magnitude limits
and galaxy types considered. The important restriction of the
morphological assessment is the uncertainty in estimating the amount
of contamination by background galaxies (e.g. Trentham \&
Tully~\cite{Trenth02}). This can lead to different authors deriving
very different slopes for the same cluster: Ferguson \&
Sandage~(\cite{Fergus88}) obtain $\alpha = -1.08 \pm 0.09$ for the
dwarf GLF in Fornax; Kambas et al.~(\cite{Kambas00}) suggest a much
steeper slope $\alpha \simeq -2.0$, based on poorer resolution data of
$2.3''$ without colour information (see also the discussion in Hilker
et al.~\cite{Hilker03}). Both surveys have comparable completeness
limits of $M_B\simeq -12$ mag. Such differences in $\alpha$ stress the
need for high-resolution imaging and an extension of the limiting
magnitude for direct cluster membership determination.

A note on nomenclature: for simplicity, throughout this paper we use
the term dE (dwarf elliptical) to refer to early-type dwarf galaxies
($M_{\rm V}>-17$ mag) in general. I.e. the term dE encompasses also dS0 and
dSph.

\subsection{Previous work by our group}
In Mieske et al.~(\cite{Mieske03}, paper I hereafter) we
investigated by means of Monte Carlo simulations the potential of the
surface brightness fluctuations (SBF) method (Tonry \&
Schneider~\cite{Tonry88}) to directly determine cluster memberships of
faint candidate dEs in nearby galaxy clusters. We find that with 1
hour $I$-band exposures on 8m class telescopes and with good seeing
(0.5$''$), reliable SBF cluster memberships can be determined down to
$M_{\rm V}\simeq -11$ mag at 20 Mpc distance. This is several magnitudes
fainter than the limit from previous spectroscopic surveys.

In Hilker et al.~(\cite{Hilker03}, paper II hereafter) we present a
wide field photometric study of the Fornax galaxy population, based on
data obtained with the WFCCD camera at the 2.5m duPont telescope at
Las Campanas Observatory, Chile (LCO). This camera has a field of view
of 25$'$ and a pixel scale of 0.8$''$, enabling the detection of very
faint dE candidates. Using a combination of visual inspection and
SExtractor (Bertin \& Arnout~\cite{Bertin96}) automated object
detection, we discovered about 70 dE candidates in Fornax with
$-12.7<M_{\rm V}<-8.5$ mag, extending the FCC sample of
Ferguson~(\cite{Fergus89}) about three magnitudes fainter.  For
constructing the GLF from the WFCCD imaging, we used the galaxies
listed in the FCC as likely cluster members plus the newly discovered
fainter dE candidates. As an ad-hoc correction for possible
interlopers, we restricted the sample to galaxies within 2$\sigma$ of
the colour-magnitude and surface brightness-magnitude relation defined
by the entire sample.  The faint end slope derived from this was
$\alpha=-1.10 \pm 0.10$, in good agreement with the value found by
Ferguson \& Sandage~(\cite{Fergus88}). Note that the error of this
value is statistical only and does not account for systematic
uncertainties of morphological classifications.

There were two deficits with the WFCCD data set, related to its large
pixel scale of 0.8$''$ and correspondingly large point source FWHM of
about 1.9$''$ ($\simeq$ 170 pc at the Fornax cluster distance): 1.
These data did not allow to measure SBF amplitudes and hence derive
direct cluster memberships for candidate dEs. 2. For $M_{\rm V}>-12$ mag,
the image resolution approached the expected size of Local Group (LG)
dE analoga, with typical LG dEs being about twice as large as the
seeing FWHM (see Fig.~\ref{mumag}). Therefore, the data did not allow
to morphologically separate probable cluster members from background
galaxies in that magnitude regime. Note that this resolution
restriction applies even more to previous surveys such as the FCC or
the Kambas et al.~(\cite{Kambas00}) work.

\subsection{Aim of this paper}
In this paper we investigate the photometric properties of the dwarf
galaxy population in the Fornax cluster ($M_{\rm V}>-17$ mag), following up
our paper II study. We use data obtained with the instrument IMACS
mounted at the 6.5m Magellan telescopes at LCO, which provides about
four times smaller pixel scale and seven times larger light collecting
area as compared to the WFCCD data of paper II, and a comparable field
of view.  In Mieske et al.~(\cite{Mieske06b}, paper III hereafter),
those data were already presented and have been used to derive a
calibration of the SBF method at blue colours.
This paper is structured as follows: 

The IMACS data are presented in
Sect.~\ref{data}. In the subsequent three chapters we describe the
following steps to compile a fiducial Fornax dE sample:

1. Confirm cluster membership of candidate dEs by means of $I$-band
SBF measurement, see Sect.~\ref{SBF}.

2. Re-assess the morphological classification of candidate dEs too
faint for SBF detection, see Sect.~\ref{morph}.

3. Search for new candidate dEs close to the resolution limit of
previous surveys, see Sect.~\ref{search}.

The photometric scaling relations and luminosity function of the
resulting fiducial Fornax dwarf galaxy population are discussed in
Sect.~\ref{revision}. In Sect.~\ref{summary} we finish this paper with
a summary and conclusions.
\section{The data}
\label{data}
\begin{figure*}
\begin{center}
  \epsfig{figure=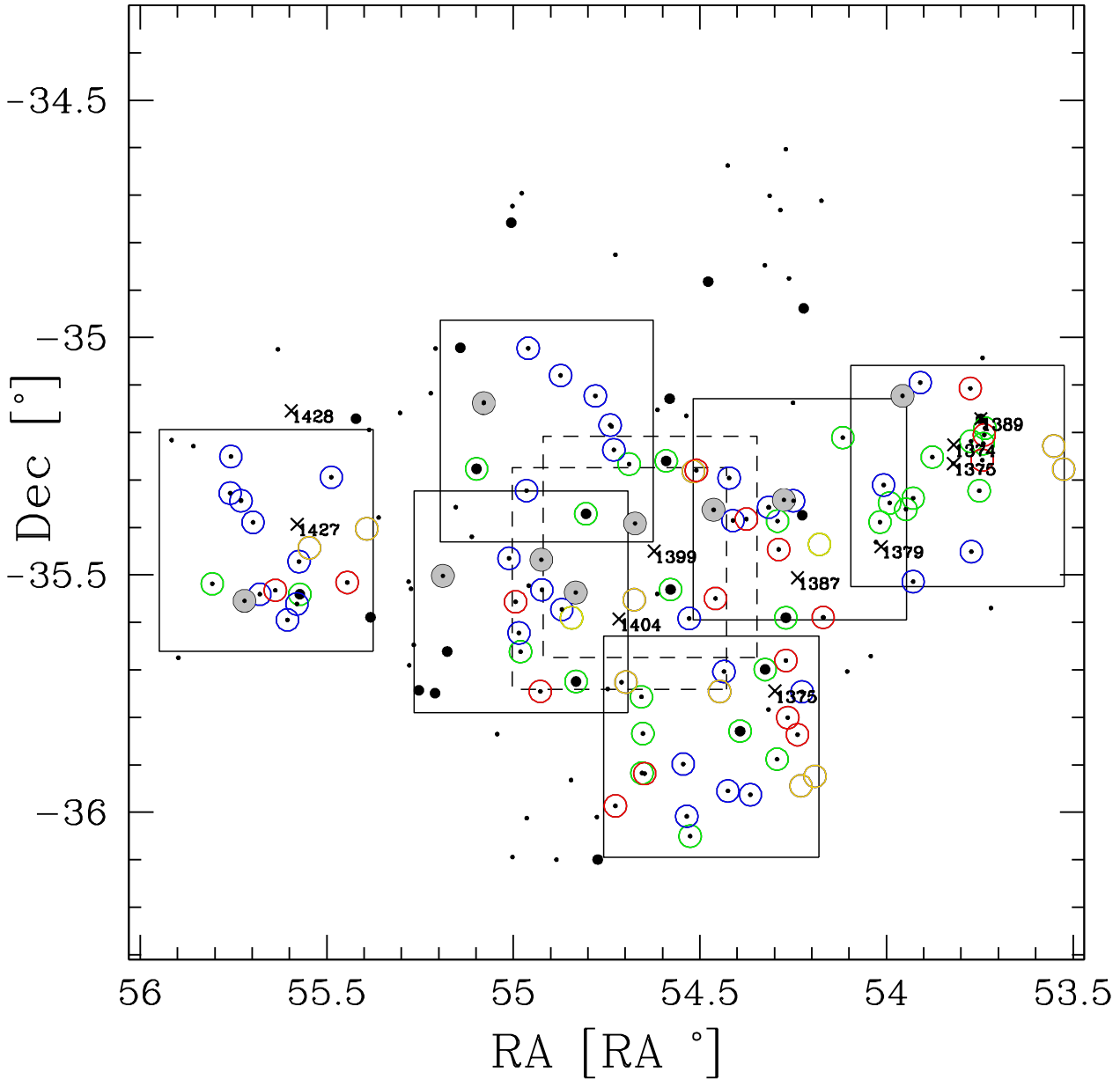,width=17.2cm}
       \caption{Map of the central Fornax cluster. Small dots are cluster member candidates from Hilker et al.~\cite{Hilker03} (paper II). Large dots are cluster members previously confirmed by radial velocity (Drinkwater et al.~\cite{Drinkw01}, Mieske et al.~\cite{Mieske04}, Hilker et al.~\cite{Hilker99}). Squares indicate the IMACS pointings from this paper. The two dashed squares indicate the two central IMACS pointings with lower integration times (see text). The coloured circles mark objects observed within those pointings. Green circles are cluster members confirmed by SBF (Sect.~\ref{SBF}). Blue circles are probable members based on a morphological analysis (Sect.~\ref{morph}). Red circles are probable background galaxies based on a morphological analysis (Sect.~\ref{morph}). Grey filled circles are unclear classifications (Sect.~\ref{morph}). Golden circles without a dot inside indicate the new dE candidates detected with the IMACS imaging (Sect.~\ref{search}). Those candidates from paper II
that were not observed were either outside the vignetted field of view or fell in the gaps between chips. Crosses mark the giant Fornax galaxies with their corresponding NGC numbers indicated.}  
\label{map}
\end{center}
\end{figure*}

The imaging data for this paper were obtained with the instrument
IMACS at Las Campanas Observatory, Chile (see also paper III). Using
the ``short'' f/2 camera, a 27.4$'$ field is imaged onto eight
2k$\times$4k chips which have a pixel scale of 0.2$''$. Seven fields
in the central Fornax cluster were observed in both $V$ and $I$ under
photometric conditions, covering the inner square degree ($\simeq$ 330
kpc) and slightly beyond (see Fig.~\ref{map}). The total $I$-band
integration time was between 3800 and 5160 seconds, except for the
central NGC 1399 field, which had 900 seconds integration for each of
the two dithered exposures. The total integration times in $V$ were
between 1200 and 1800 seconds except for the central field around NGC
1399, for which it was 900 seconds for each of the two slightly
dithered exposures. The main part of the observing time was used for
the $I$-band in order to measure SBF in this filter.
Table~\ref{sbfresults} lists the candidate dEs from paper II that were
imaged with IMACS.

The image reduction before SBF measurement was done in the following
steps: first, a master-bias was created for each chip and was
subtracted from the domeflat exposures. Then for each chip the bias
corrected dome flats were combined to create master-domeflats. Having
the master-biases and master-domeflats prepared for each chip, we used
the COSMOS package\footnote{Carnegie Observatories System for
  MultiObject Spectroscopy,
  http://llama.lco.cl/$\sim$aoemler/COSMOS.html} to do
bias-subtraction, trimming and flat-field correction of the raw
science frames. The reduced single science frames were registered with
integer pixel shifts to avoid distortion of the SBF power spectrum and
combined using a min-max rejection algorithm. The seeing FWHM of the
combined images typically ranged between 0.6 and 1.2$''$, with a
median around 0.8$''$.
\begin{figure}
\begin{center}
  \epsfig{figure=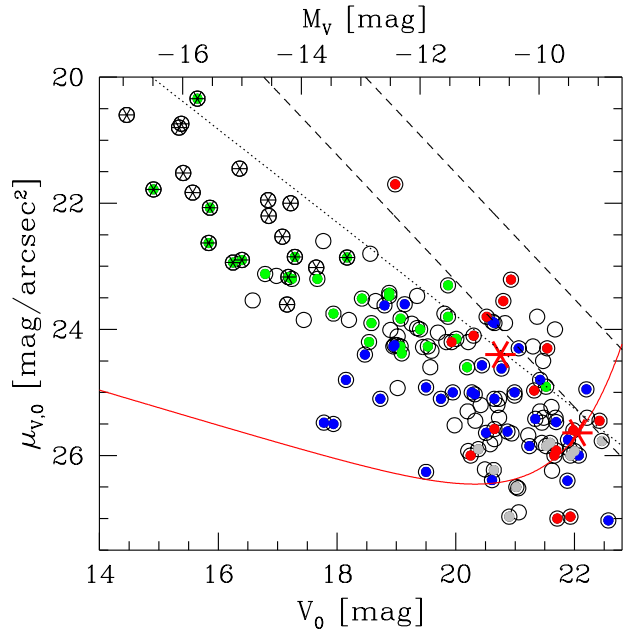,width=8.6cm}
       \caption{Central surface brightness $\mu_{\rm V,0}$ plotted vs. $V_0$ of the candidate dEs from Hilker et al.~\cite{Hilker03} (paper II), based on WFCCD data. The solid line indicates the 50\% completeness limit. The two large red asterisks indicate two candidate dEs found in a background WFCCD field. Compared to the average candidate density in the Fornax fields, their presence implies a formal contamination fraction between 10\% and 50\%. Small black asterisks indicate galaxies with confirmed cluster membership from radial velocity (Drinkwater et al.~\cite{Drinkw01}, Mieske et al.~\cite{Mieske04}, Hilker et al.~\cite{Hilker99}). Coloured circles indicate objects re-observed with the higher resolution IMACS data presented in this paper.  See Fig.~\ref{map} for a map of the imaged Fornax cluster region. Green filled circles indicate galaxies with cluster membership confirmed from SBF (Sect.~\ref{SBF}). Blue filled circles indicate galaxies with probable cluster membership based on revised morphology assessment (Sect.~\ref{morph}). Red filled circles indicate probable background galaxies based on revised morphological assessment (Sect.~\ref{morph}). {\bf Grey filled circles} indicate unclear classifications (Sect.~\ref{morph}). The dotted line is a fit to the $M_{\rm V}$ - $\mu_{\rm V,0}$ values of Local Group dEs (Grebel et al. \cite{Grebel03}). The lower dashed line indicates the location of an exponential galaxy light profile with FWHM=3.5$''$, which is twice the seeing for the paper II data. The upper dashed line indicates an exponential profile with FWHM=1.5$''$, which is about twice the seeing in the IMACS data presented in this paper.}  
\label{mumag}
\end{center}
\end{figure}

\begin{figure}[ht!]
\begin{center}
  \epsfig{figure=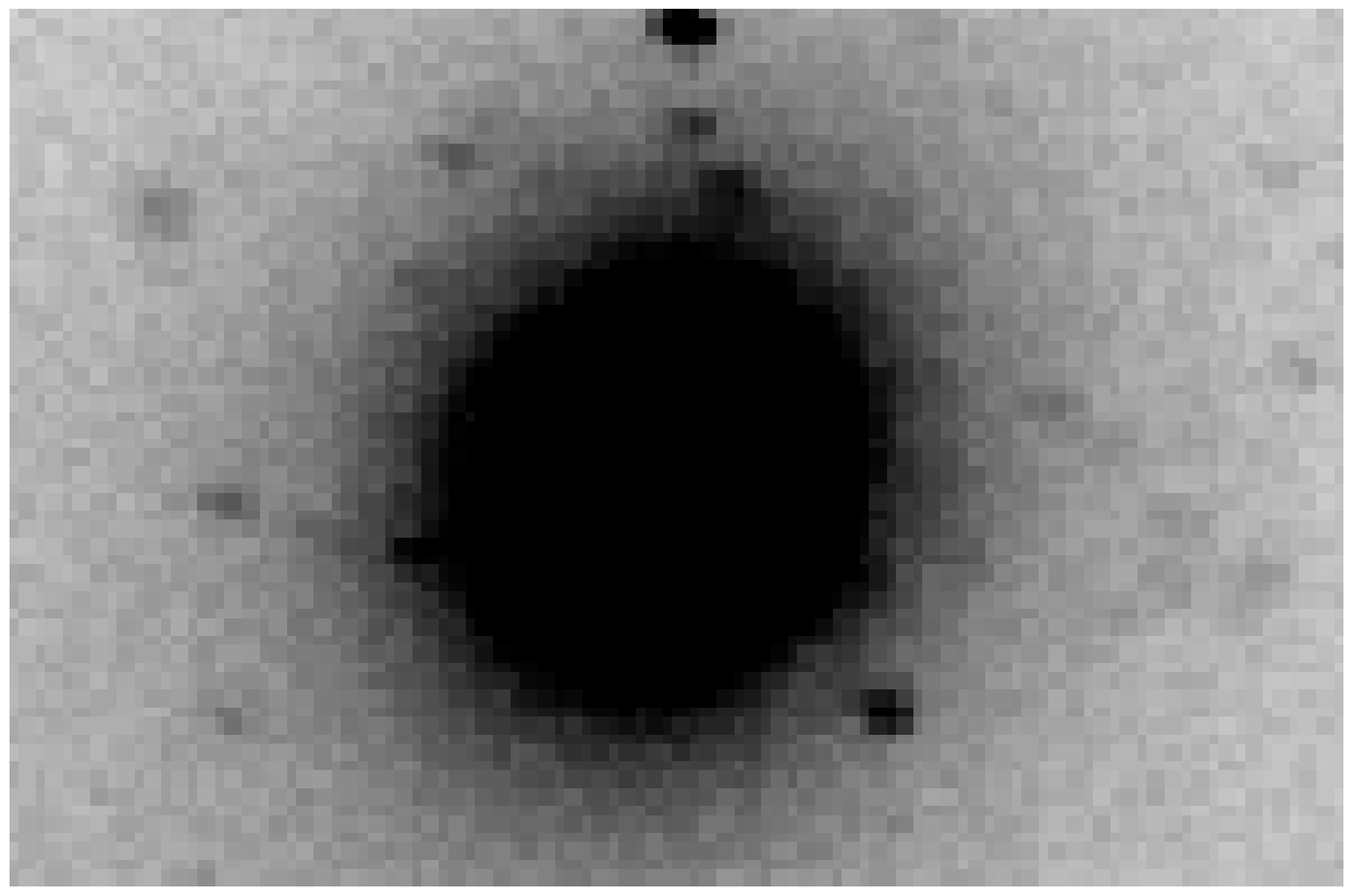,width=2.75cm}\hspace{0.2cm}
  \epsfig{figure=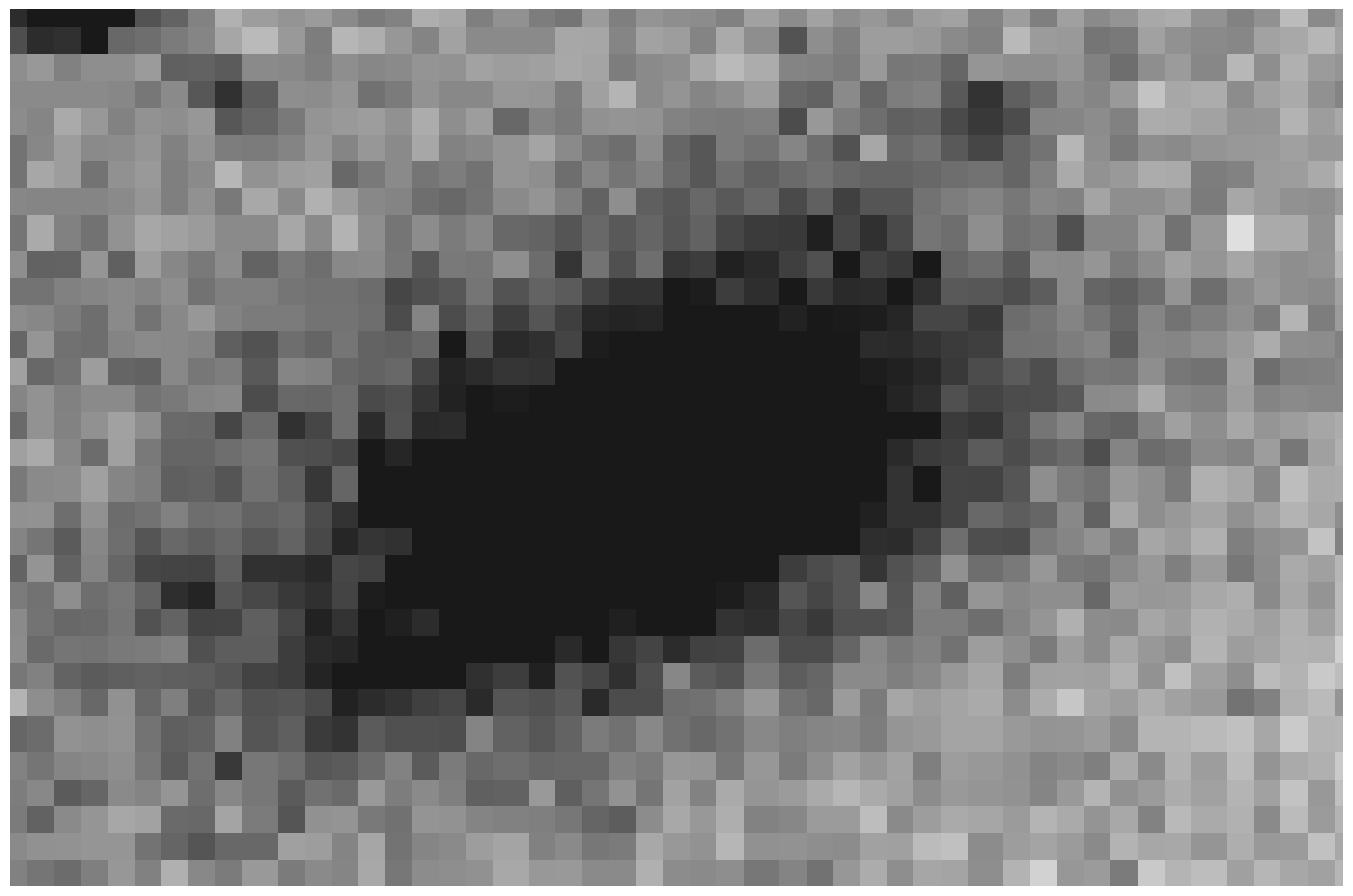,width=2.75cm}\hspace{0.2cm}
  \epsfig{figure=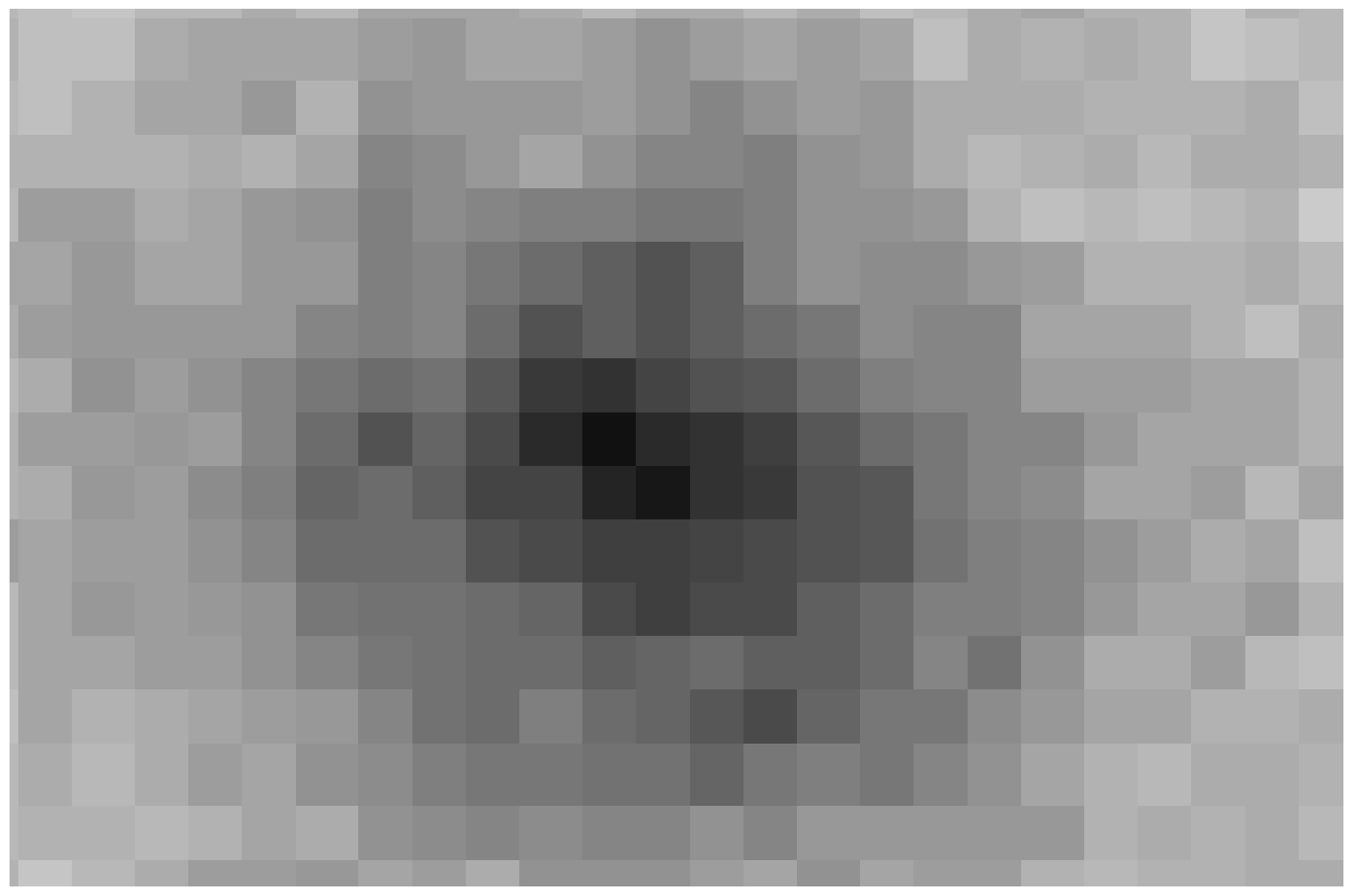,width=2.75cm}\vspace{1cm}\\
  \epsfig{figure=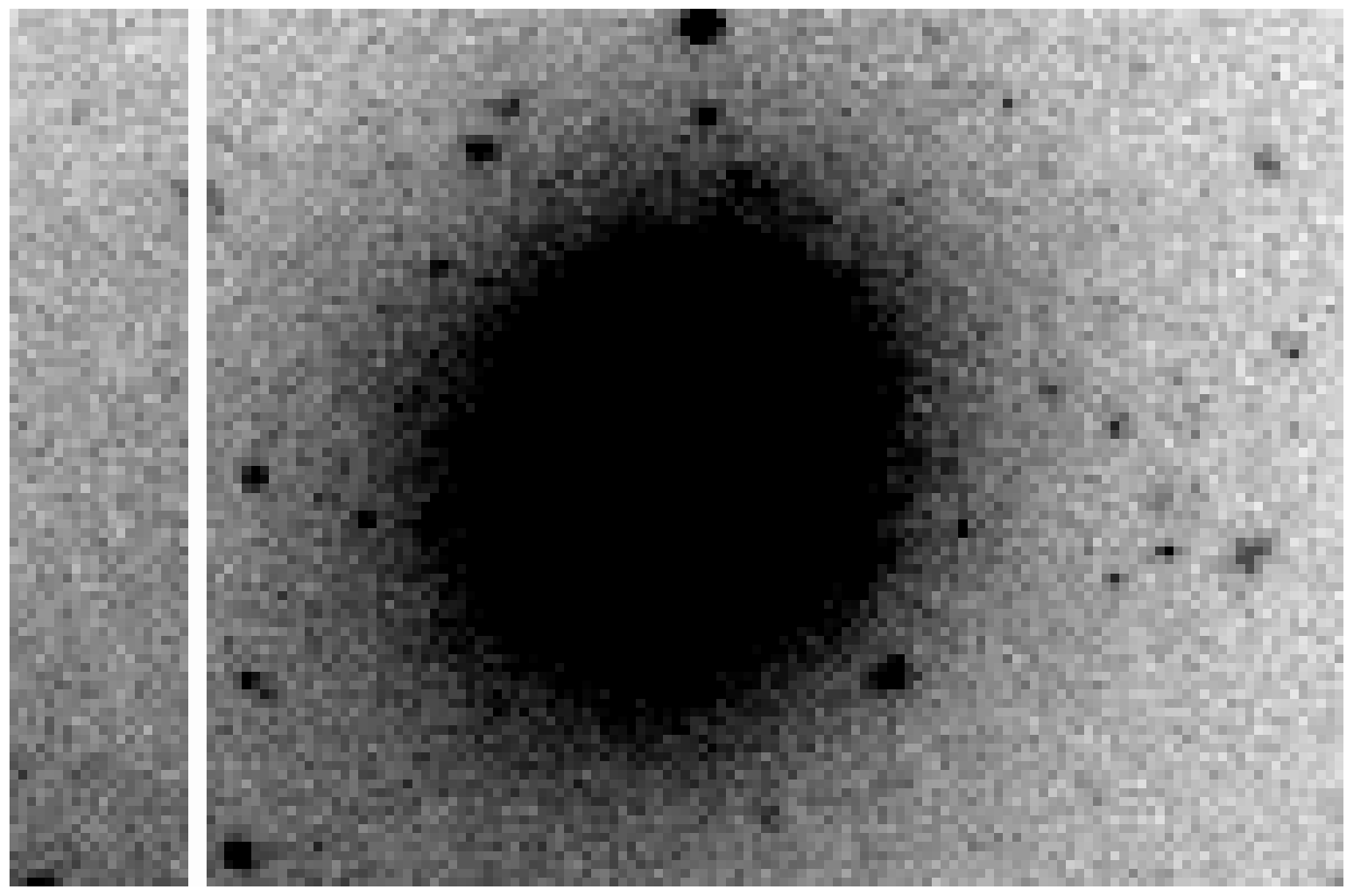,width=2.75cm}\hspace{-2.75cm}\parbox[t]{2.75cm}{\vspace{-2.5cm}\hspace{0.5cm}\Large \bf FCC 222}\hspace{0.2cm}
  \epsfig{figure=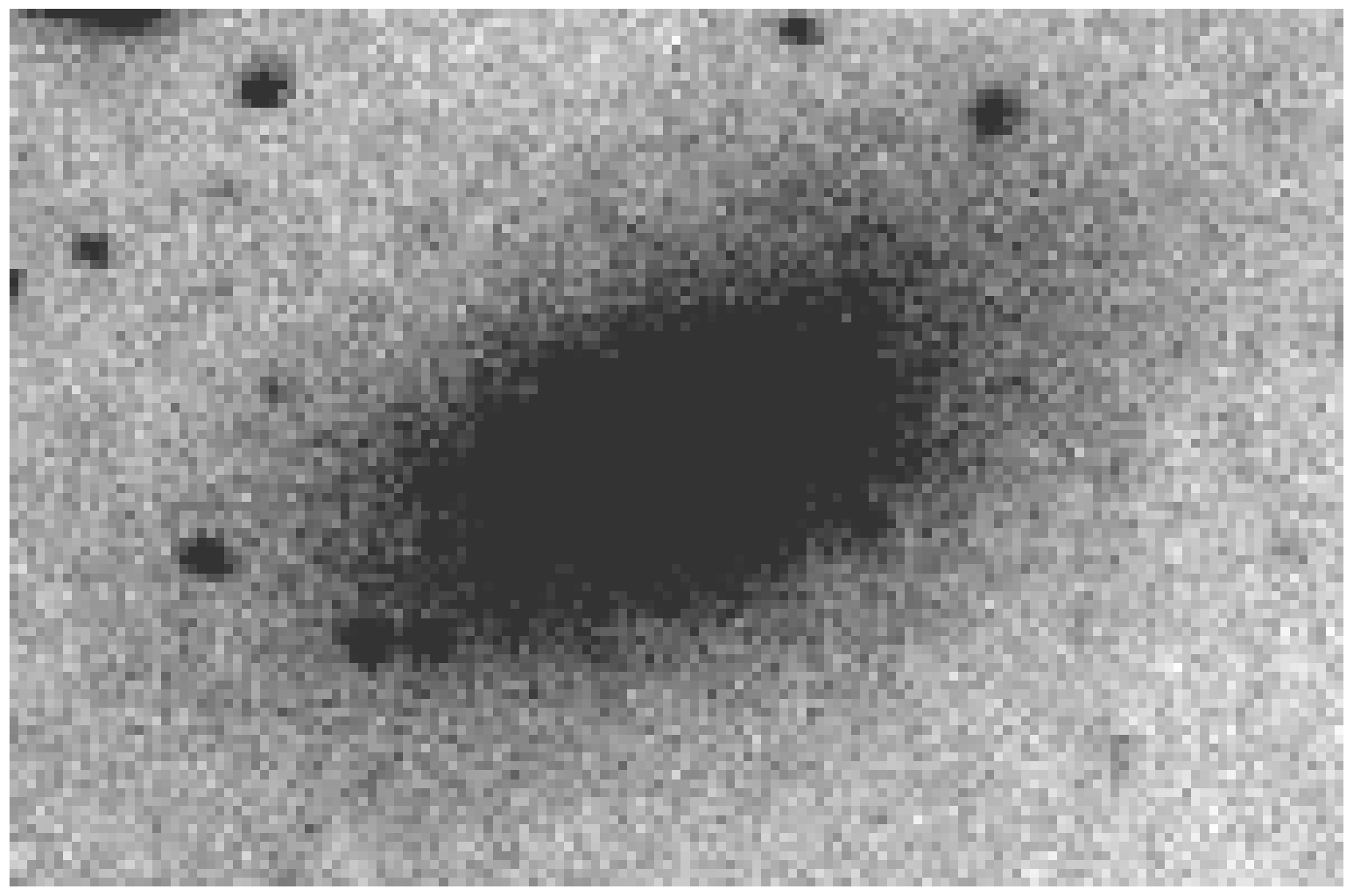,width=2.75cm}\hspace{-2.75cm}\parbox[t]{2.75cm}{\vspace{-2.5cm}\hspace{0.5cm}\Large \bf FCC 196}\hspace{0.2cm}
  \epsfig{figure=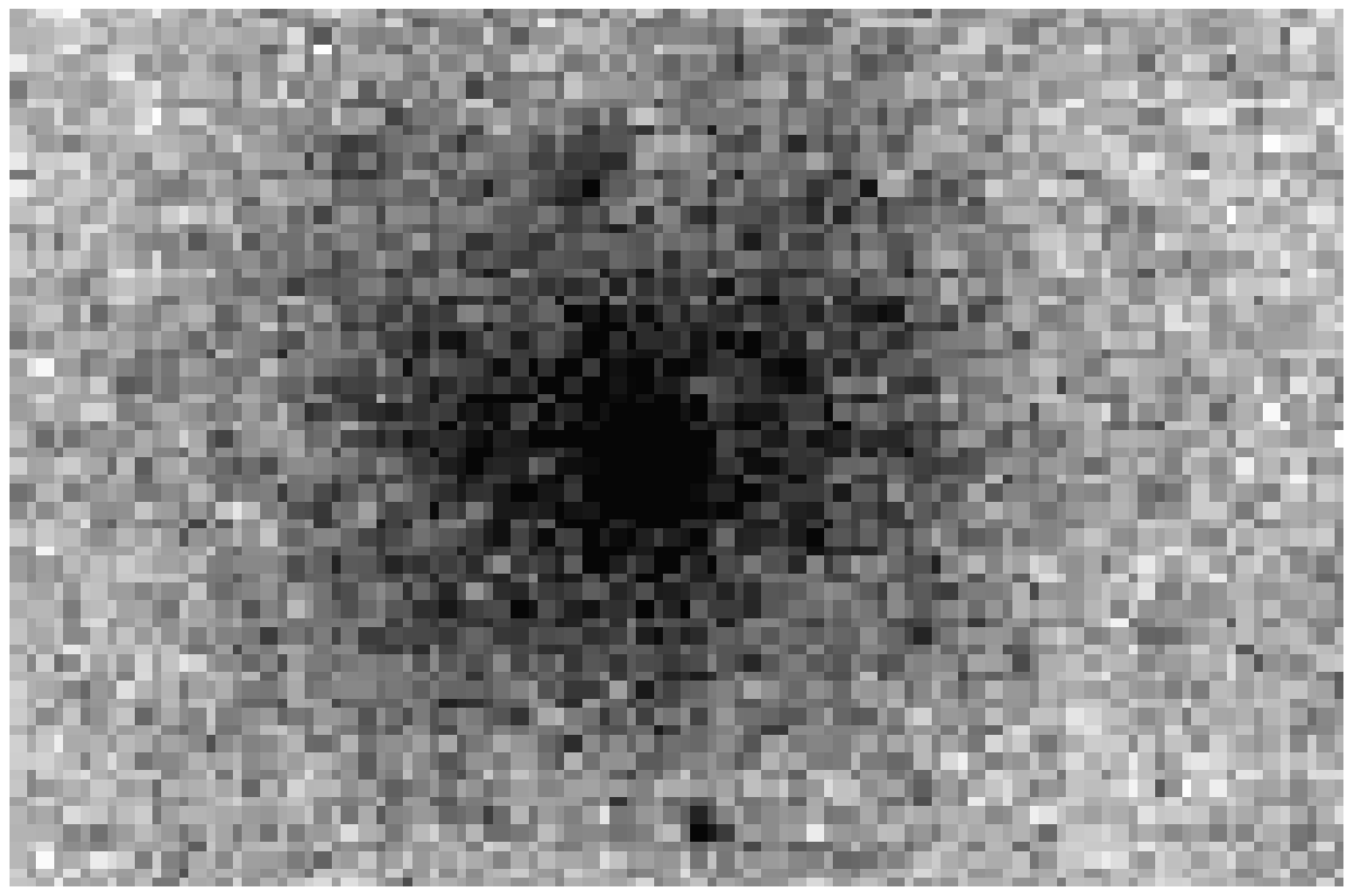,width=2.75cm}\hspace{-2.75cm}\parbox[t]{2.75cm}{\vspace{-2.5cm}\hspace{0.5cm}\Large \bf FCC 191}
\caption{Thumbnails of SBF confirmed Fornax cluster members (Sect.~\ref{SBF}), i.e. those with flag=1 in Table~\ref{sbfresults}. {\bf Top:} WFCCD images. The PSF FWHM is about 1.8$''$. {\bf Bottom:} IMACS images. The PSF FWHM is about 0.8$''$. The thumbnail sizes are from left to right 77$\times$50$''$ (7$\times$4.6 kpc at the Fornax cluster distance), 38$\times$25$''$ (3.5$\times$2.3 kpc), and 19$\times$12$''$ (1.8$\times$1.1 kpc).}  
\label{memSBF}
\end{center}

\end{figure}
\begin{figure}[ht!]
\begin{center}
  \epsfig{figure=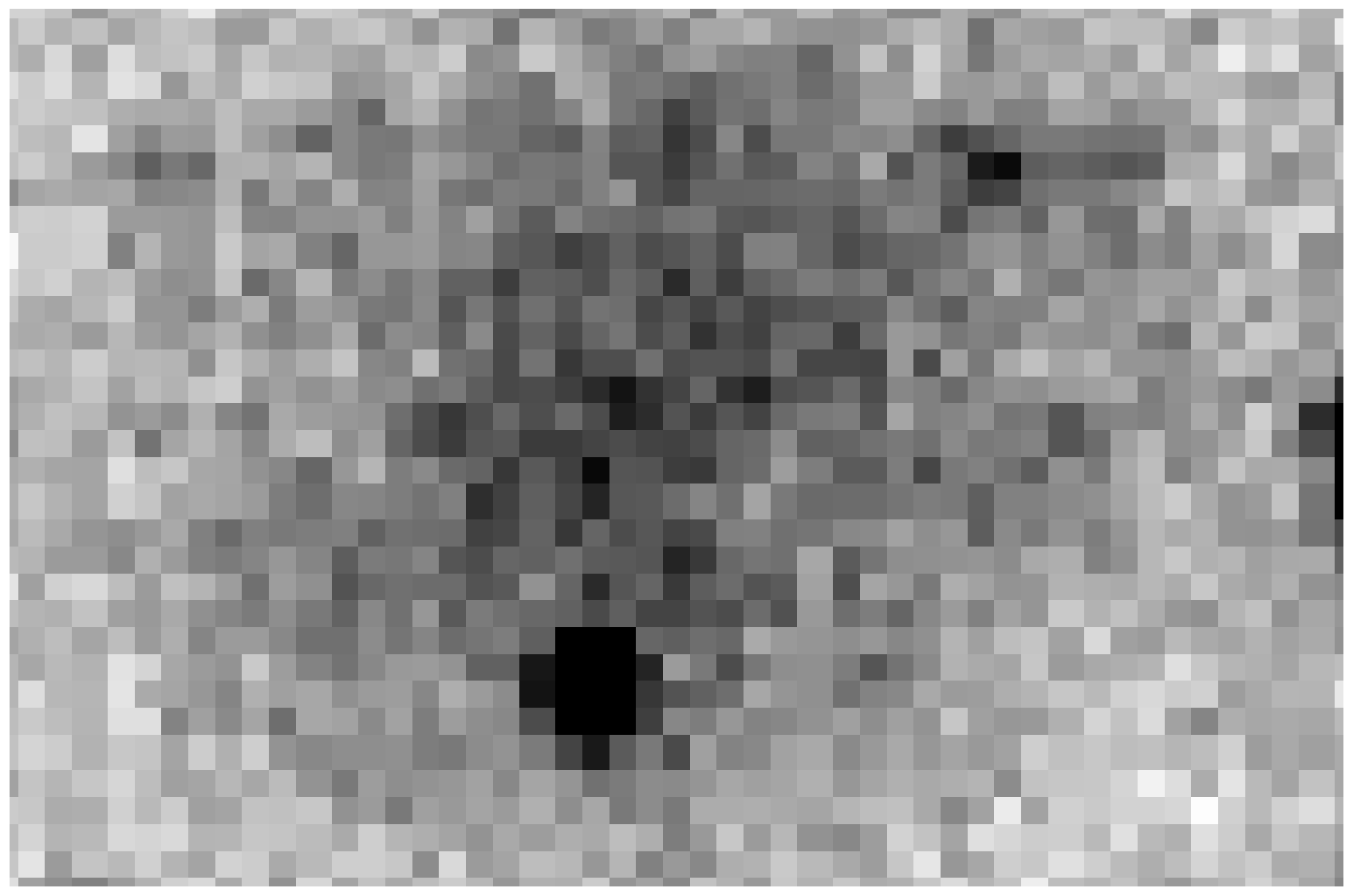,width=2.75cm}\hspace{0.2cm}
  \epsfig{figure=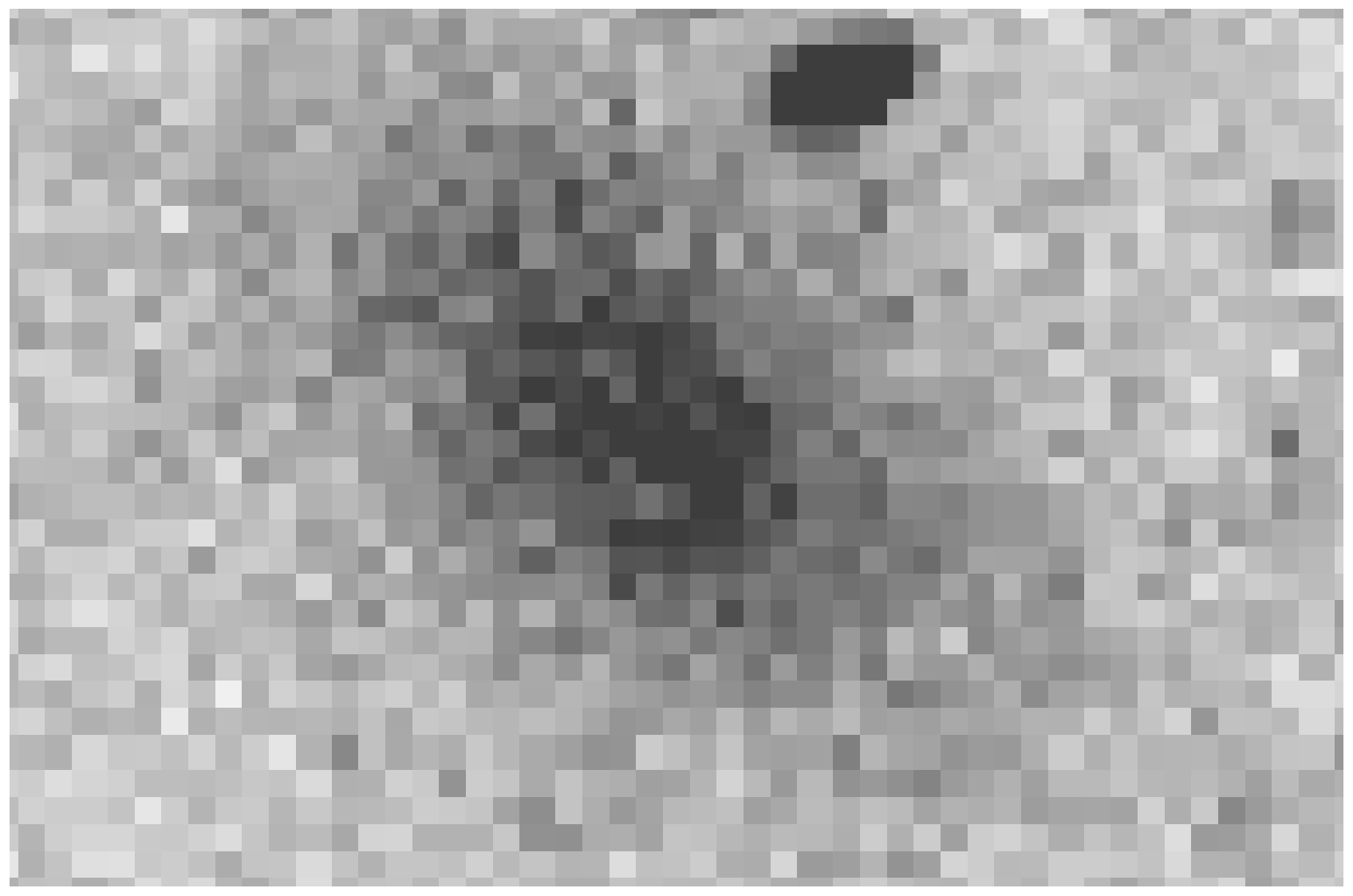,width=2.75cm}\hspace{0.2cm}
  \epsfig{figure=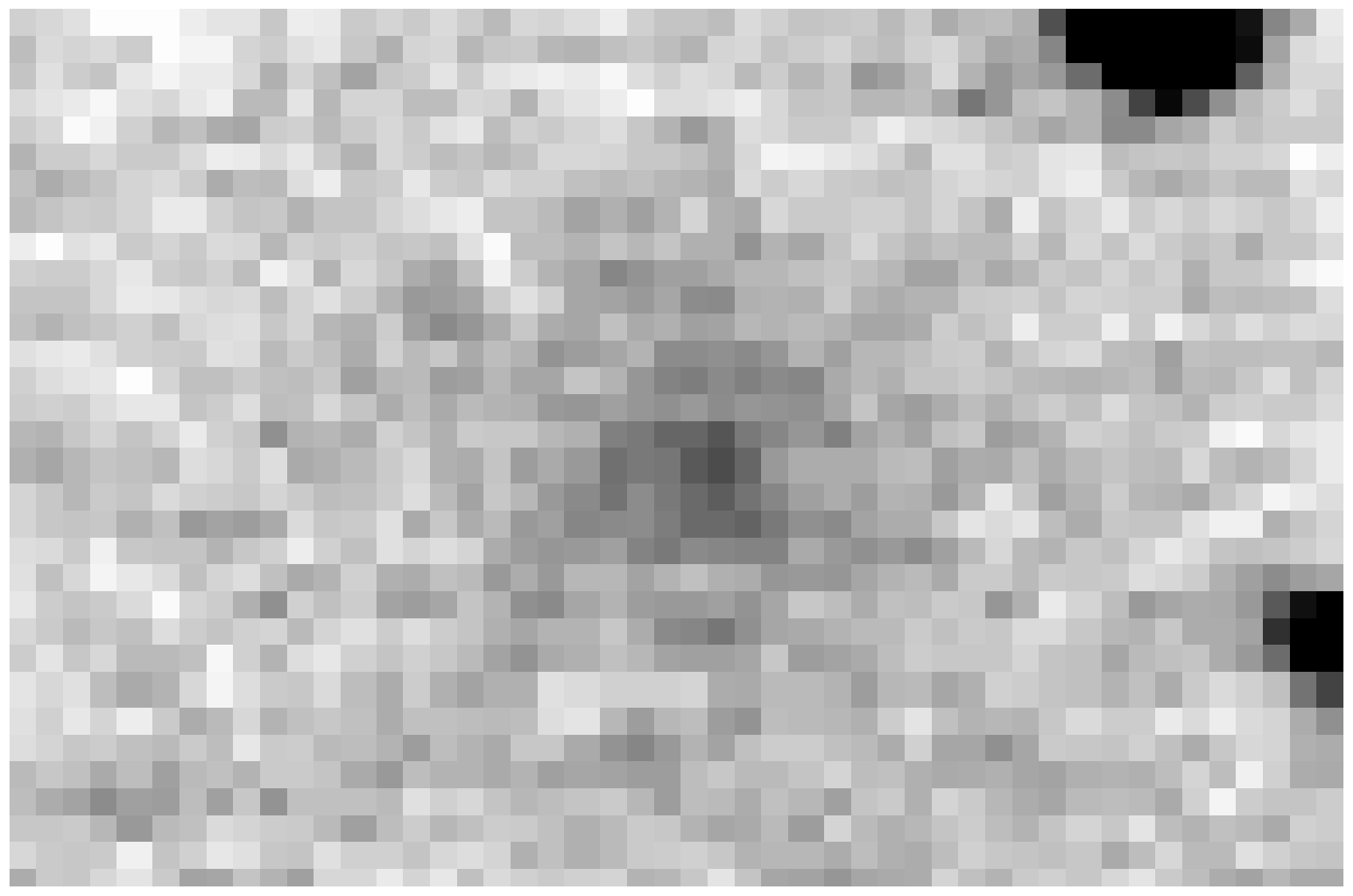,width=2.75cm}\vspace{1cm}\\
  \epsfig{figure=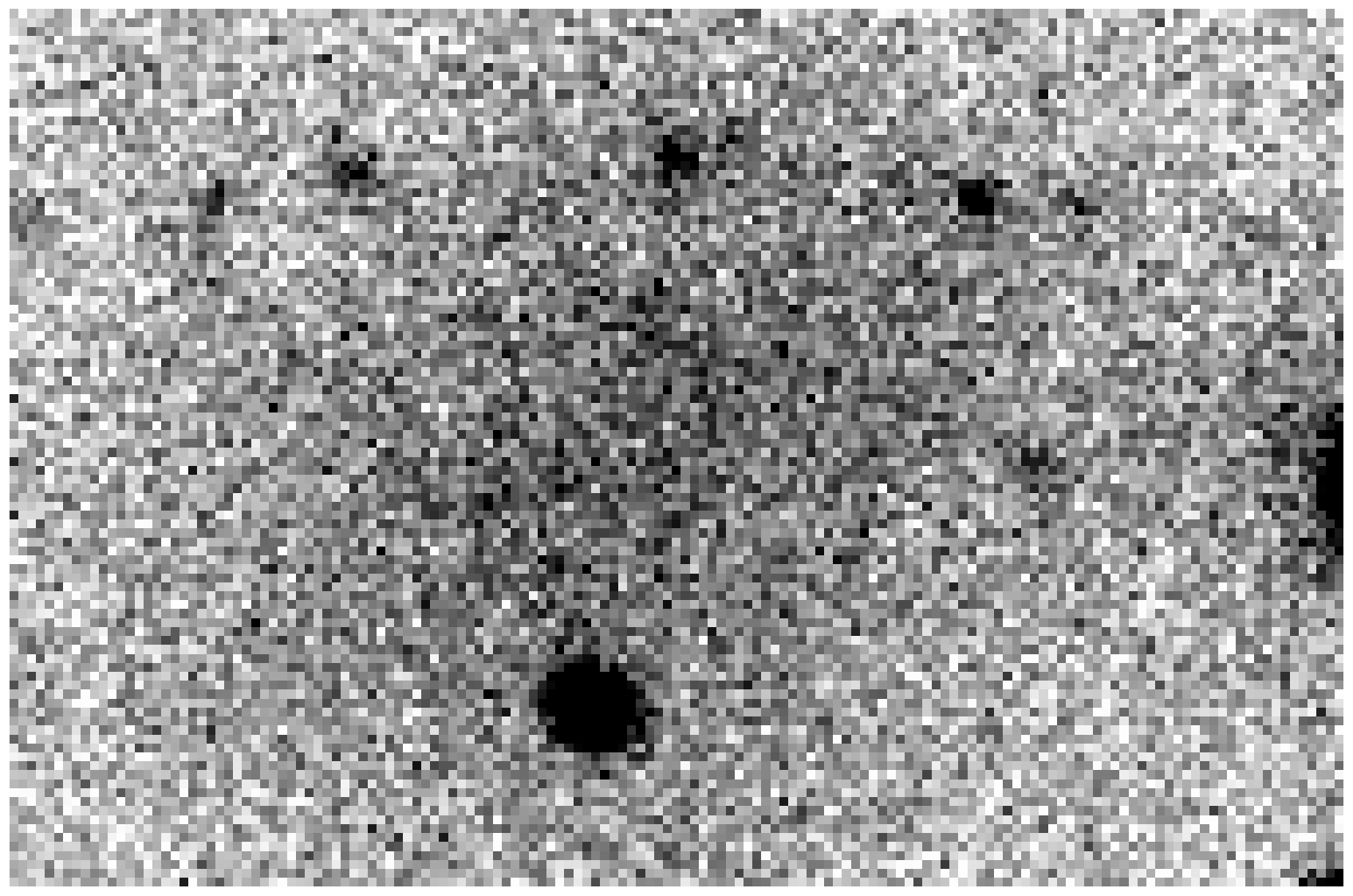,width=2.75cm}\hspace{-2.75cm}\parbox[t]{2.75cm}{\vspace{-2.5cm}\hspace{0.1cm}\Large \bf   FCC 269}\hspace{0.2cm}
  \epsfig{figure=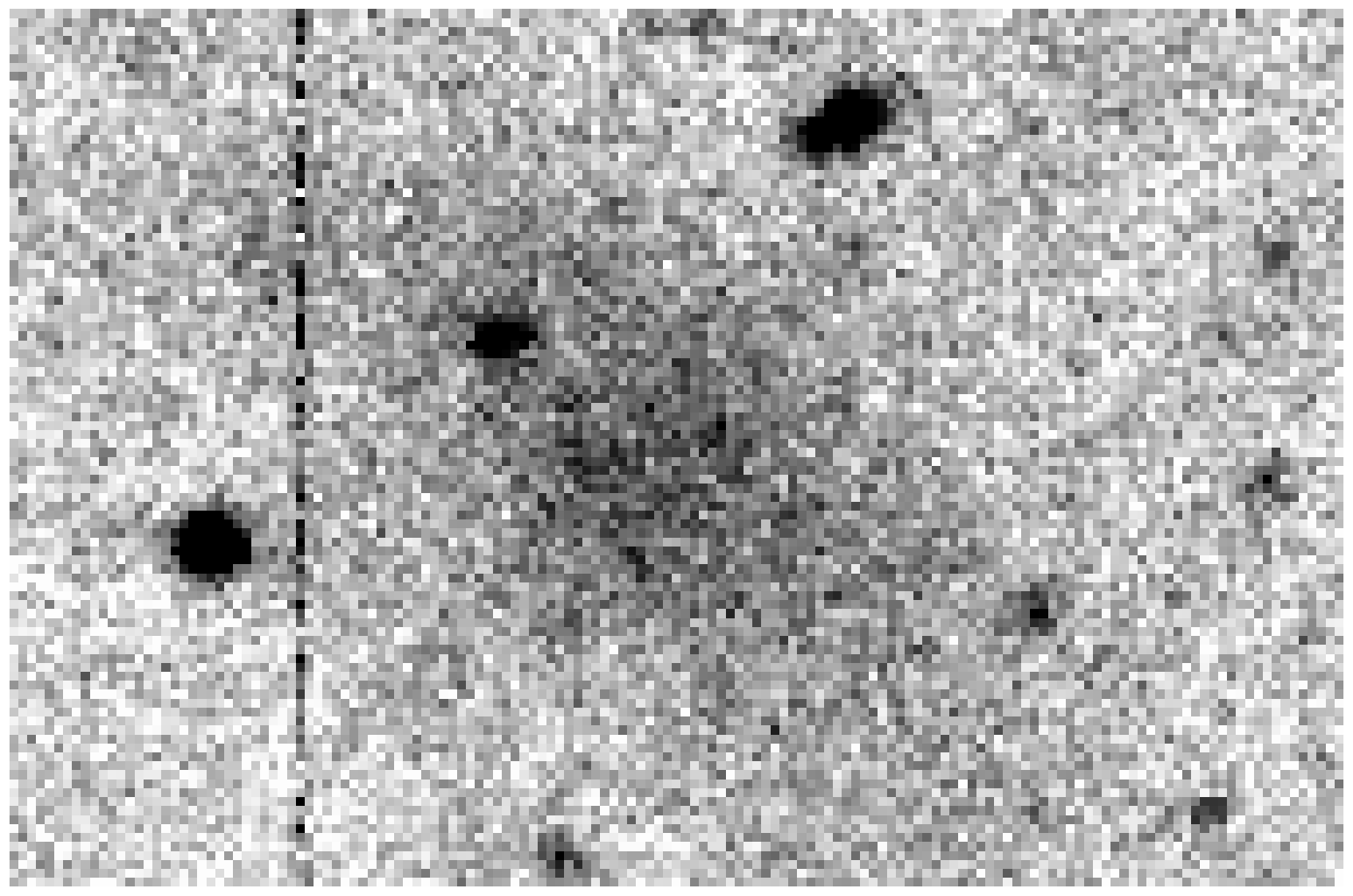,width=2.75cm}\hspace{-2.75cm}\parbox[t]{2.75cm}{\vspace{-2.5cm}\hspace{0.1cm}\Large \bf   FCC 284}\hspace{0.2cm}
  \epsfig{figure=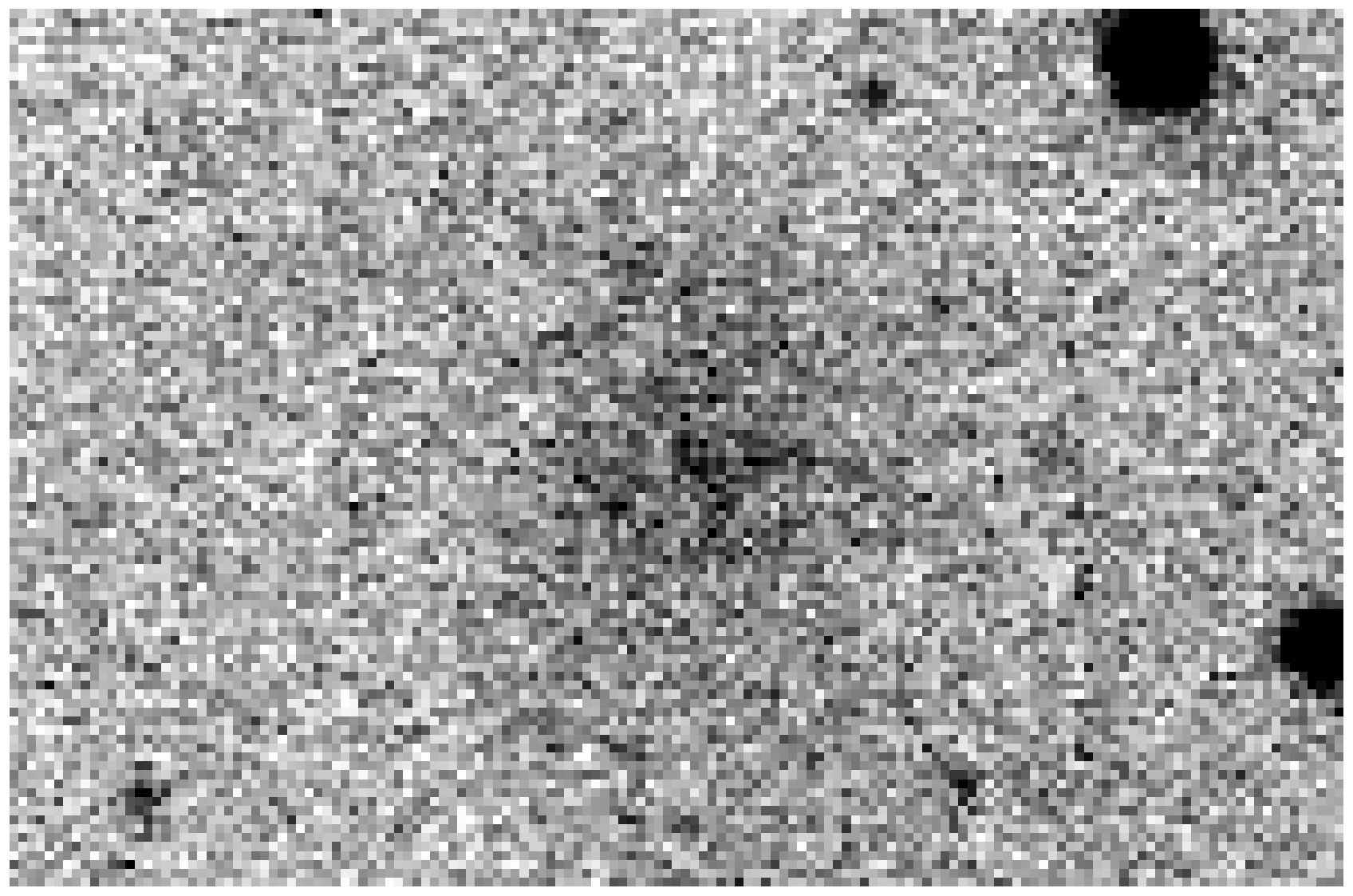,width=2.75cm}\hspace{-2.75cm}\parbox[t]{2.75cm}{\vspace{-2.5cm}\hspace{0.1cm}\Large \bf WFLSB 2-1}
\caption{Thumbnails of morphologically confirmed Fornax cluster members (Sect.~\ref{morph}), i.e. those with flag=2 in Table~\ref{sbfresults}. {\bf Top:} WFCCD images. {\bf Bottom:} IMACS images. The thumbnail sizes are 38$\times$25$''$ (3.5$\times$2.3 kpc at the Fornax cluster distance).}  
\label{memmorph}
\end{center}

\end{figure}
\begin{figure}[ht!]
\begin{center}
  \epsfig{figure=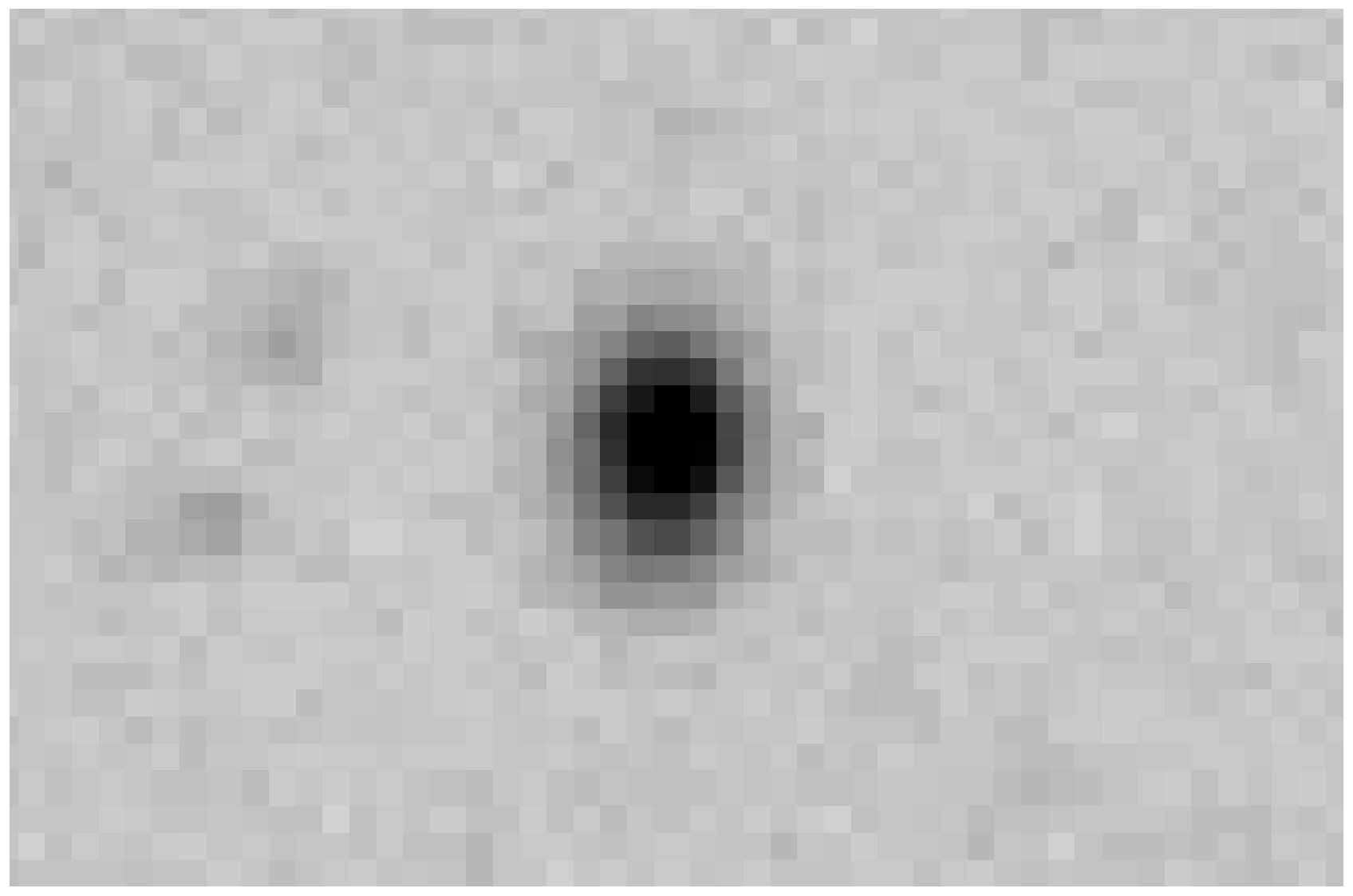,width=2.75cm}\hspace{0.2cm}
  \epsfig{figure=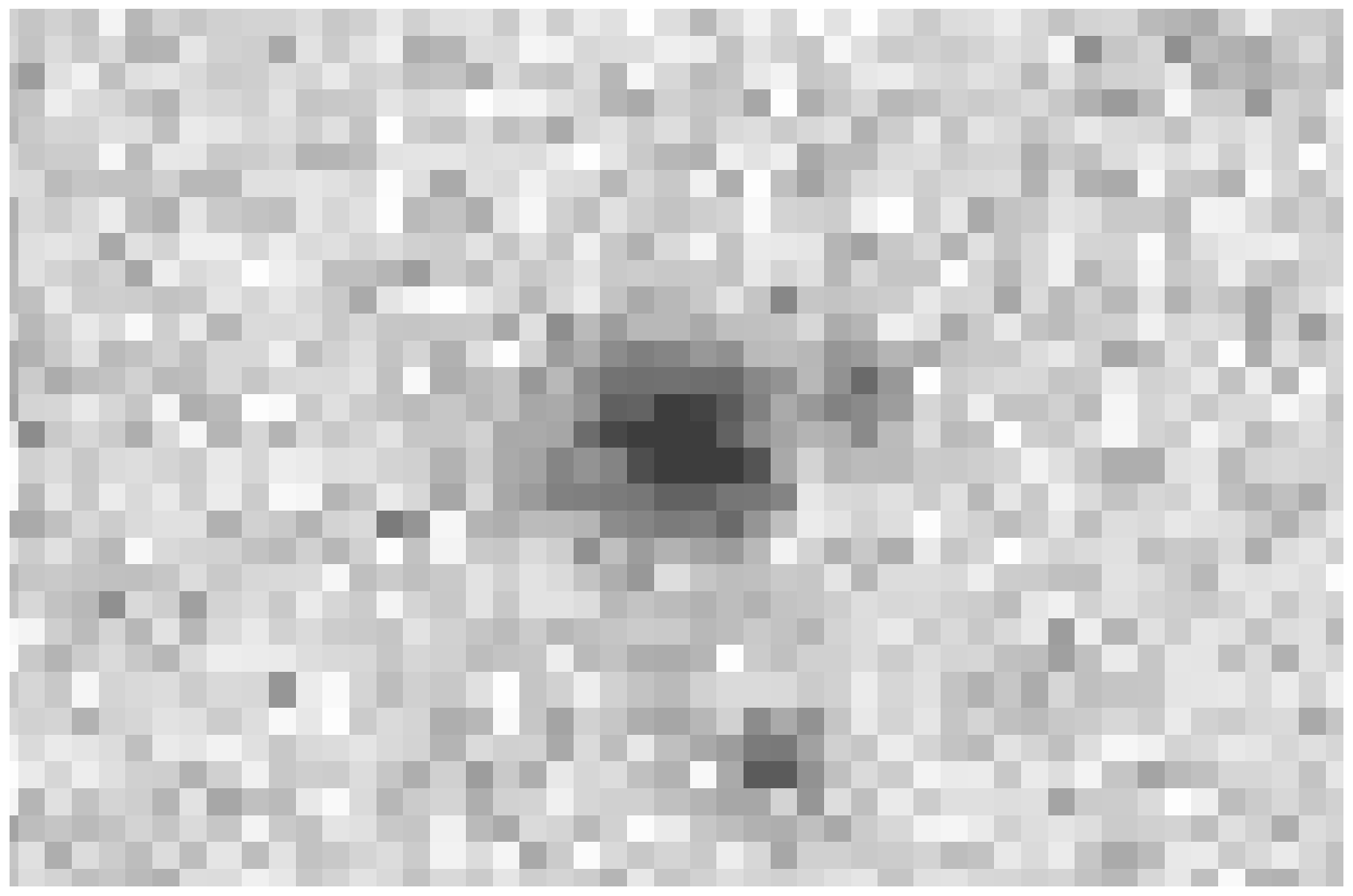,width=2.75cm}\hspace{0.2cm}
  \epsfig{figure=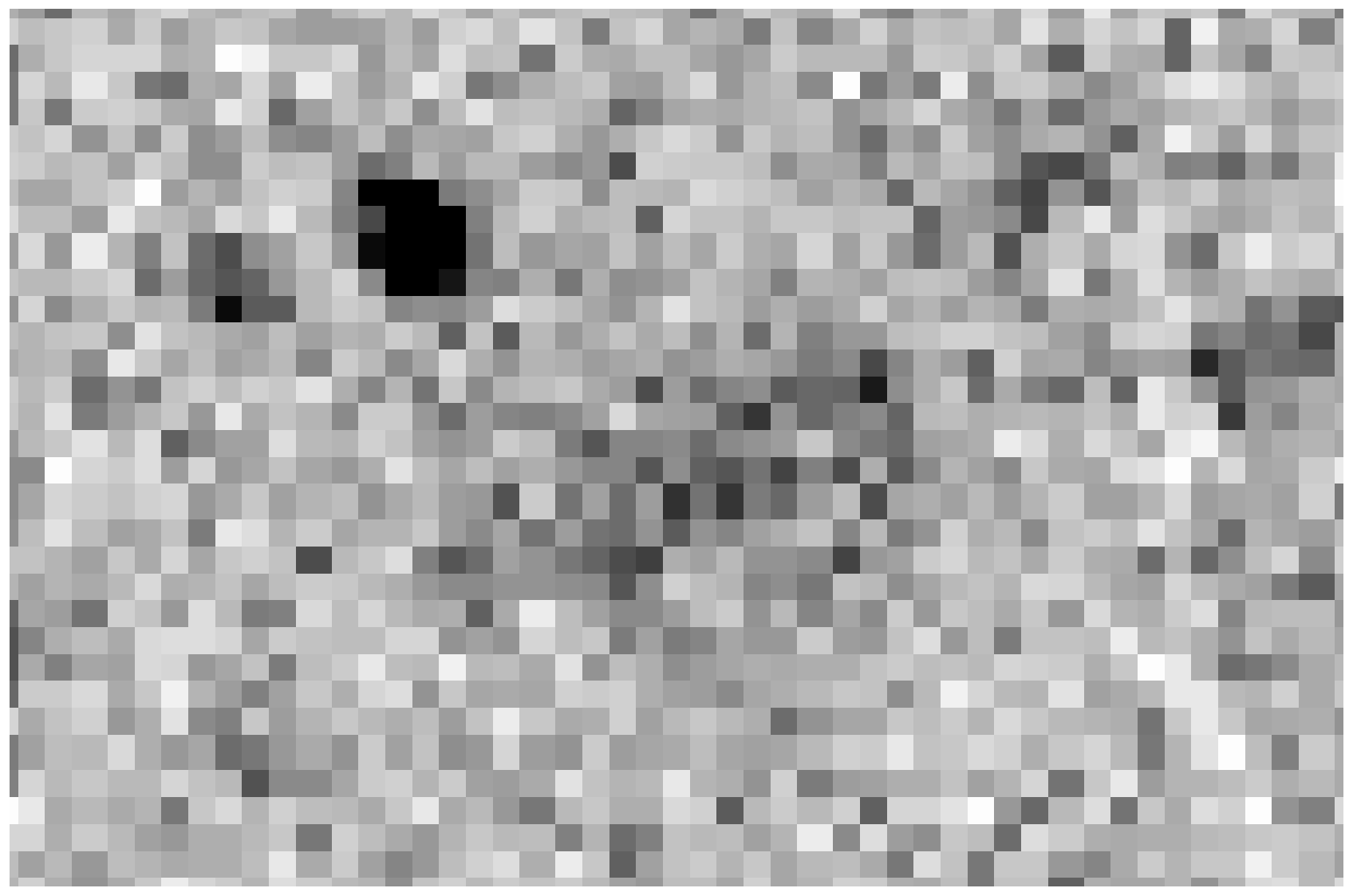,width=2.75cm}\vspace{1cm}\\
  \epsfig{figure=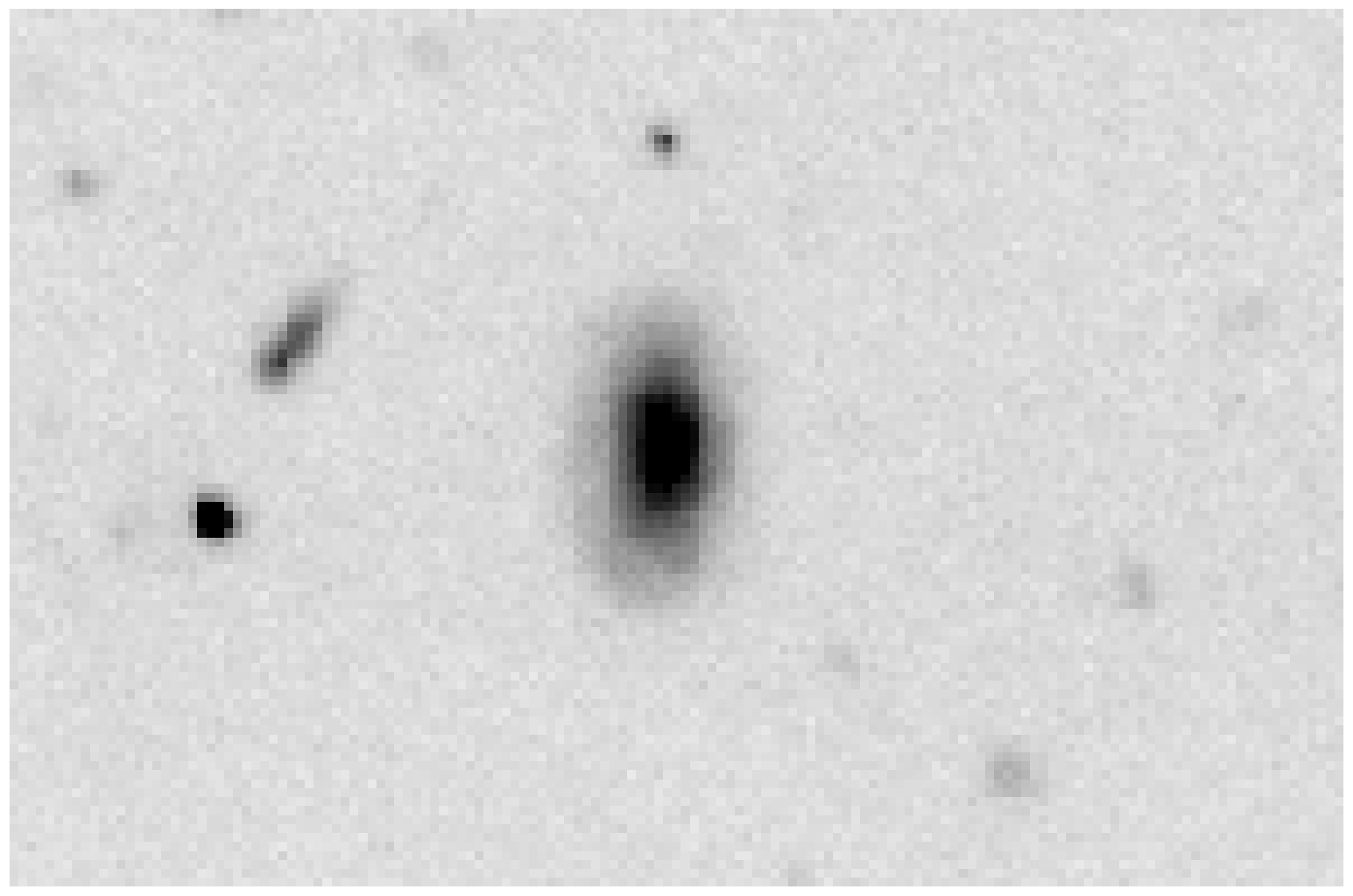,width=2.75cm}\hspace{-2.75cm}\parbox[t]{2.75cm}{\vspace{-2.5cm}\hspace{0.1cm}\Large \bf   FCC 141}\hspace{0.2cm}
  \epsfig{figure=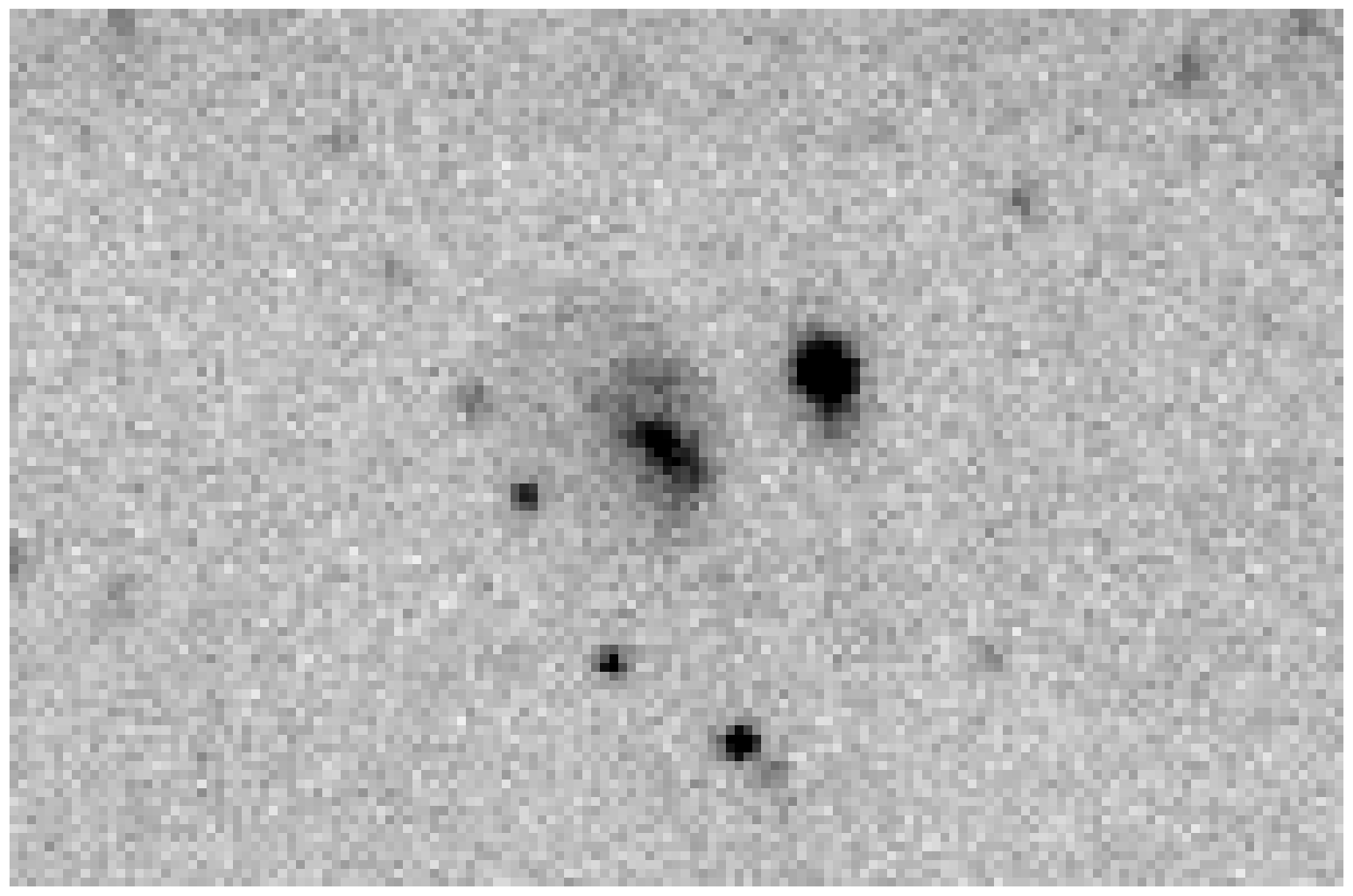,width=2.75cm}\hspace{-2.75cm}\parbox[t]{2.75cm}{\vspace{-2.5cm}\hspace{0.1cm}\Large \bf   WFLSB 6-3}\hspace{0.2cm}
  \epsfig{figure=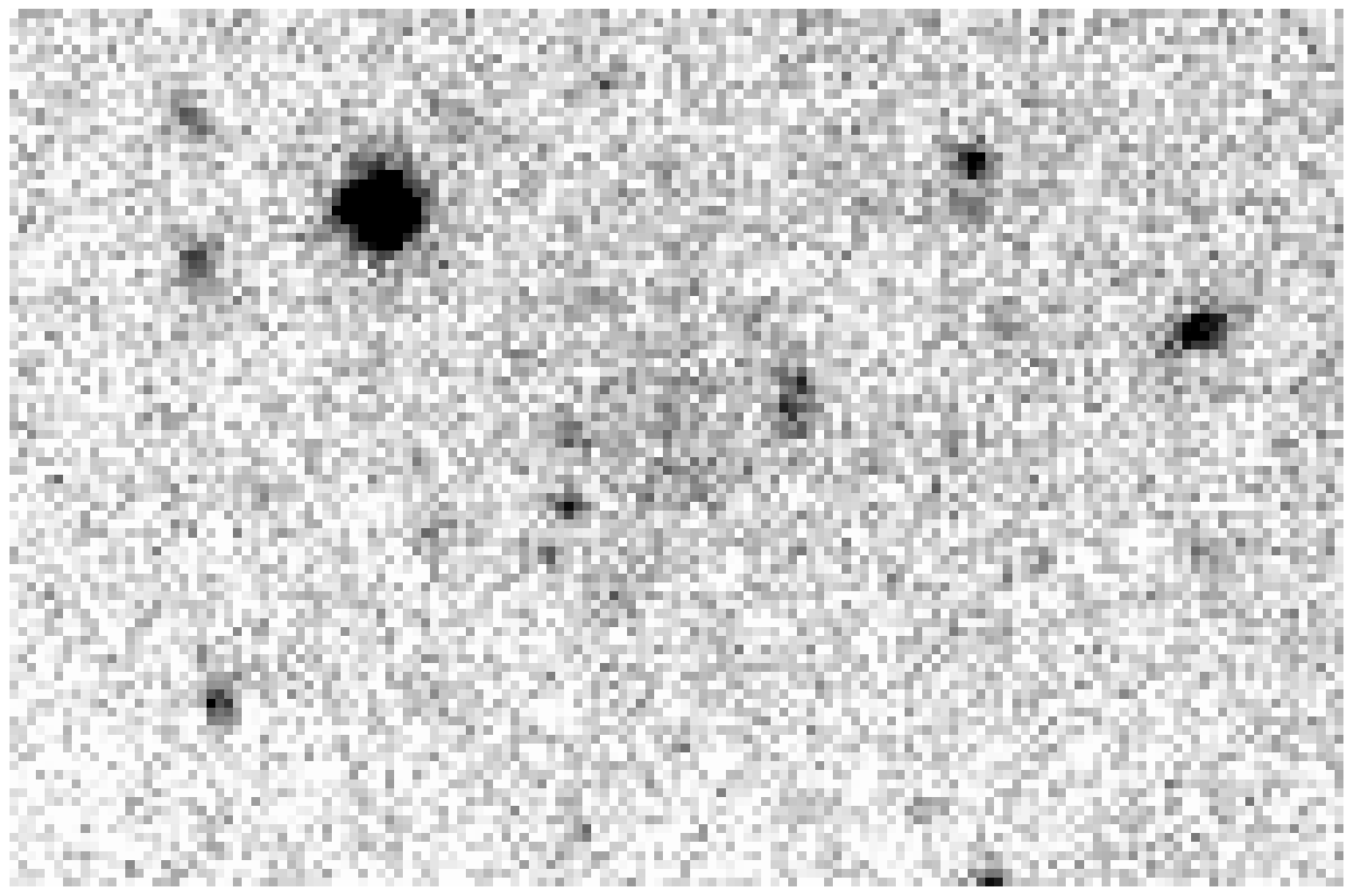,width=2.75cm}\hspace{-2.75cm}\parbox[t]{2.75cm}{\vspace{-2.5cm}\hspace{0.1cm}\Large \bf   WFLSB 1-4}
\caption{Left two thumbnails: two morphologically re-classified background galaxies (Sect.~\ref{morph}), i.e. those with flag=3 in Table~\ref{sbfresults}. Right thumbnail: a source with unclear classification, i.e. those with flag=4 in Table~\ref{sbfresults}. {\bf Top:} WFCCD images. {\bf Bottom:} IMACS images. The thumbnail sizes are 38$\times$25$''$ (3.5$\times$2.3 kpc at the Fornax cluster distance).}  
\label{backmorph}
\end{center}

\end{figure}

\section{Direct cluster membership assignment via SBF measurement}
\label{SBF}
The SBF measurements for the candidate dEs are performed in the
$I$-band and are described in detail in paper III. The signal-to-noise
ratio of the SBF measurement is defined as $S/N=\frac{P_0}{P_1}$,
where $P_0$ is the amplitude of the proper stellar SBF signal, while
$P_1$ is the white noise amplitude. The amplitude of the background
SBF fluctuations (arising from undetected intra-cluster globular
clusters, background galaxies, and CCD effects like fringing) ranged
between 0.1 and 0.55 mag.

We classified as confirmed cluster members all galaxies for which
  the SBF measurement had a $S/N$ $>3$, see Table~1 of paper III. All
  those 25 galaxies have SBF distances compatible to within
  1.9$\sigma$ with the Fornax cluster reference distance of 31.39
  $\pm$ 0.12 mag when applying the calibration D from paper III. This
  specific calibration implies a bifurcated relation between colour
  $(V-I)_{\rm 0}$ and absolute fluctuation magnitude
  $\overline{M}_{\rm I}$. In paper III it was marginally favoured over
  a non-bifurcated relation at the 1.8$\sigma$ level. When opting for
  a non-bifurcated relation instead, the scatter of the implied SBF
  distances naturally increases (see Tables~2 and 3 of paper III). In
  that case, one galaxy is attributed a very low distance
  more than 3$\sigma$ below the reference distance, namely FCC 218
  with $(m-M)=$30.32 $\pm$ 0.33 mag. For its magnitude, this galaxy is
  not an outlier in terms of colour or surface brightness, see
  Table~\ref{sbfresults} and Sect.~\ref{revision}. Ex- or including it
  in the sample of Fornax cluster members does therefore not influence
  the statements that will be made further on in this paper.
  
  The 25 galaxies with SBF confirmed cluster membership span the range
  $-16.6<M_{\rm V}<-11.2$ mag.  Those galaxies are assigned a
membership flag 1 in Table~\ref{sbfresults} of this paper (see
  also Fig.~\ref{memSBF} for example thumbnail images). There are 7
additional galaxies for which an SBF signal with $S/N$ $<$ 3 was
detected. Three of them have background SBF below 50\% of the stellar
SBF signal and an SBF distance error below 0.5 mag. We include those
three sources into the sample of confirmed cluster members assigning a
flag value of 1.3. This yields a full sample of 28 galaxies spanning a
luminosity range $-16.6<M_{\rm V}<-10.1$ mag. The remaining four dEs
($-12.4<M_{\rm V}<-10.4$ mag) with larger background fluctuations and
SBF distance errors were defined as probable members from morphology
with a membership flag of 1.7 (see also next Section).  Note that nine
of the 28 SBF confirmed members also have radial velocity measurements
available (Drinkwater et al.~\cite{Drinkw01}), all of which are
consistent with them being cluster members.
\section{Morphological re-classification}
\label{morph}

In addition to allowing for SBF measurements, the IMACS data also
enable us to check which dE candidates from paper II retain a smooth
morphology when imaged at 2-3 times better spatial resolution.

For this morphological re-assessment, we defined three possible cases.
i) The galaxy retains its smoothness on the IMACS data, hence is
confirmed as probable cluster member (flag 2 in
Table~\ref{sbfresults}). ii) The galaxy exhibits clear substructure
that is indicative of a background spiral, or resolves into separate
point sources (flag 3 in Table~\ref{sbfresults}). Note that for the
latter case we also demand that there is no low surface brightness
envelope. Such an envelope would be indicative of a dwarf irregular
galaxy (dIrr) that may be in the cluster. We did in any case not
detect any such possible dIrr in our image inspections. iii) The IMACS
data do not allow a clear assessment (flag 4 in
Table~\ref{sbfresults}). The latter case iii) is restricted to very
low surface brightness candidates (Fig.~\ref{mumag}).
Figs.~\ref{memmorph} and~\ref{backmorph} show example cases for the
morphological re-assessment. In the magnitude-surface brightness plot
in Fig.~\ref{mumag} we indicate the respective classifications of the
re-observed objects by different colour codings. We note that this
morphological classification implicitly exploits the fact that the
dwarf galaxy population in the central Fornax cluster is vastly
dominated by early-type galaxies (Ferguson~\cite{Fergus89}).

The majority of candidate dEs is morphologically confirmed as probable
cluster members: out of 60 candidate dEs with revised morphological
assessment (and no SBF measurement) in the range $-13.2<M_{\rm
  V}<-8.8$ mag, 34 are confirmed as probable members, 17 are
re-classified as probable background, and 9 have unclear definitions.
The contamination by background galaxies hence is about 1/3 (17 out of
51). However, for the calculation of the GLF in paper II we excluded
galaxies more than 2$\sigma$ outside the mean surface-brightness
magnitude relation to reduce the sample contamination. When
considering only candidates within the 2$\sigma$ surface brightness
limit and brighter than the 50\% detection limit of $M_{\rm V}=-9.8$
mag from paper II, the contamination drops to about 20\%.
Fig.~\ref{mumag} shows that as expected, the sources re-classified as
background galaxies lie close to the resolution limit of the WFCCD
data.  We finally note that the WFCCD data from paper II included one
pointing towards an empty comparison field outside the Fornax cluster
in which two candidate dEs were detected (see Fig.~\ref{mumag}). Given
that the average number of candidate dEs per pointing in Fornax was
6.5 $\pm$ 0.7 (paper II), the formal background contamination derived
from the detection of these two sources is in the range 10\% to 50\%.
This is consistent with the amount of contamination estimated from the
IMACS morphological re-assessment.  \vspace{0.5cm}

Apart from morphological re-classification, we have also re-measured
surface brightness profiles, total magnitudes and colours for the
paper II candidate dEs. We describe these measurements in detail in
Sect.~\ref{revision}, but mention this re-measurement here since in
the next subsection we refer to the revised surface
brightness-magnitude plot in Fig.~\ref{mumag_imacs}.
\subsection{SBF background galaxies?}
\label{sbfback}
\begin{figure}
\begin{center}
  \epsfig{figure=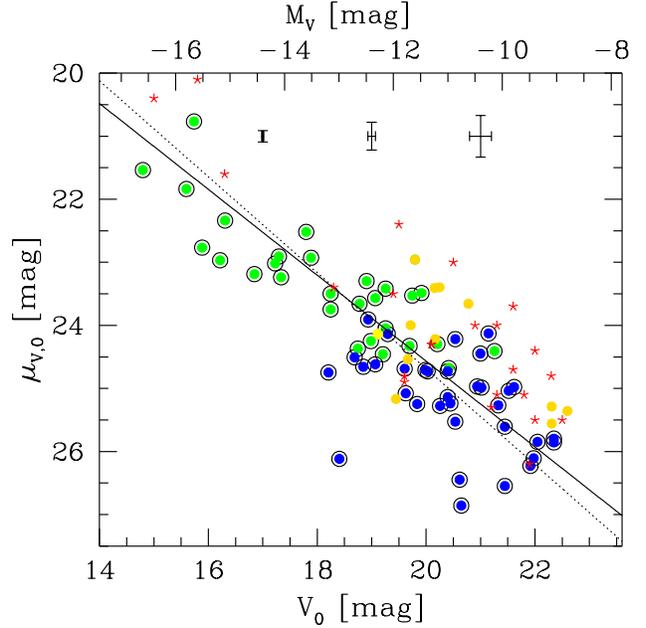,width=8.6cm}
       \caption{Magnitude-surface brightness plot as in Fig.~\ref{mumag},
now using the new photometry of the IMACS data and plotting only probable cluster members from SBF memberships and morphology (Sects.~\ref{SBF} and~\ref{morph}). Typical error bars are indicated The magnitudes were redenning corrected using Schlegel et al.~(\cite{Schleg98}). The golden dots indicate newly found dE candidates in the IMACS data (Sect.~\ref{search}). The solid line is a fit to the data points applying a 3$\sigma$ clipping. The dotted line shows the fit from the WFCCD photometry of all dE candidates  from Hilker et al.~(\cite{Hilker03}, paper II), also applying a 3$\sigma$ clipping. The red asterisks indicate Local Group dEs (Grebel et al.~\cite{Grebel03}). }  
\label{mumag_imacs}
\end{center}
\end{figure}

Do some galaxies in our sample classify as SBF confirmed background
galaxies? This would be the case for galaxies without an SBF signal
that are bright and large enough to expect such a signal at the Fornax
distance. In that context it is striking that all morphologically
selected cluster members for which no SBF signal was detected had
faint surface brightnesses $\mu_{\rm V,0} \ge$ 24 mag / arcsec$^2$ (see
Fig.~\ref{mumag_imacs}). There is only a small overlap region in
surface brightness ($24<\mu_{\rm V,0}<24.6$ mag/arcsec$^2$) where both SBF
memberships and morphological memberships are assigned. Galaxies in
this overlap region with no SBF signal detected are the only ones that
could in principle be confirmed as SBF background galaxies. Of course,
one also expects such an overlap even if all galaxies are cluster
members. This is because of differing SBF detection limits among the
galaxies, depending on observing conditions (seeing, integration time)
and intrinsic SBF amplitudes.  The overall surface brightness limit
for SBF detection in our data is $\simeq 24.3$ mag / arcsec$^2$
(except for the central pointing with shorter integration time). This
limit is between the estimates derived in paper I for seeing FWHM of
0.5$''$ and 1.0$''$ and at comparable integration times, which fits to
the fact that the median seeing of our data was 0.8$''$.

In the following, we briefly discuss the galaxies in the overlap
surface brightness region. The detectability of the SBF signal
decreases at a fixed surface brightness when going to galaxies with
fainter total luminosities, given that the SBF sampling area
decreases. This fact can explain the occurence of 3 galaxies with just
morphological membership in the range $20.5<V<21.5$ mag,
$24<\mu_{\rm V,0}<24.6$ mag (Fig.~\ref{mumag_imacs}).  We find two further
galaxies in the same $\mu_{\rm V,0}$ range but about 1.5 mag brighter. These
are indeed the only two sources which from this plot may qualify as
SBF confirmed background galaxies.  The brighter one of these galaxies
is FCC 197. This galaxy actually had a SBF signal consistent with the
cluster distance, but it is one of those four sources whose $S/N$ was
too low and background fluctuation too large to reliably classify it
as an SBF member (flag=1.7 in Table~\ref{sbfresults}). The fainter
galaxy is FCC 220. For this galaxy we did not detect a measurable SBF
signal. The corresponding galaxy image has comparably bad seeing
($\simeq 0.9 ``$), and the galaxy is quite red ($(V-I)\simeq 1.14$).
These two facts decrease the detectability of the SBF signal. FCC 220
is furthermore only a few tenths of magnitudes brighter than the
approximate limiting surface brightness for SBF detection in
Fig.~\ref{mumag_imacs}.

We therefore conclude that for none of the galaxies, the lack of a
detectable SBF signal is sufficient to classify them as background.
\begin{figure}
\begin{center}
  \epsfig{figure=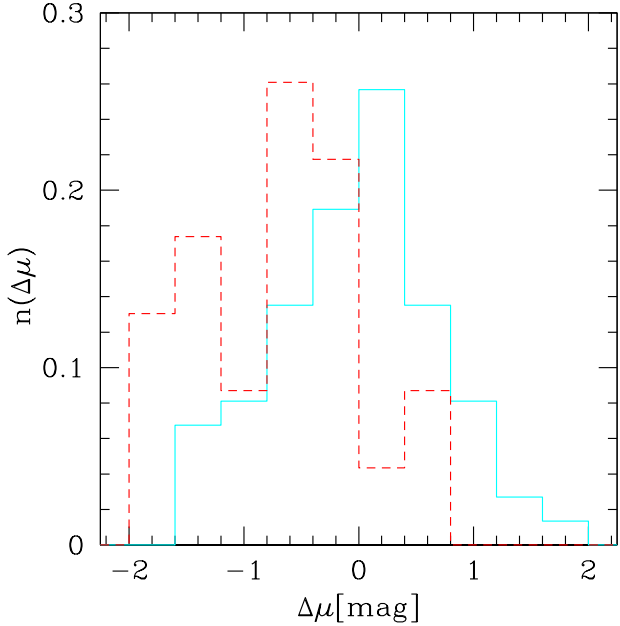,width=8.6cm}
       \caption{This plot shows the normalised distribution of central surface brightnesses of the Fornax sample (solid histogram) and Local Group sample (dashed histogram) {\it relative to} the surface-brightness magnitude relation in Fornax (Eq.~\ref{mumagrel}). By definition, the Fornax distribution is centered on 0. This plot shows that the overall shift in surface brightness between Fornax and Local Group is due to {\it both} a lack of small and overabundance of large galaxies in Fornax. According to a KS test, both distributions share the same parent distribution at only 0.08\% probability. When shifting the Local Group distribution such that the mean of both distributions agree, the KS probability is 99.9\%. }  
\label{deltamu}
\end{center}
\end{figure}
\begin{figure*}[ht!]
\begin{center}
  \epsfig{figure=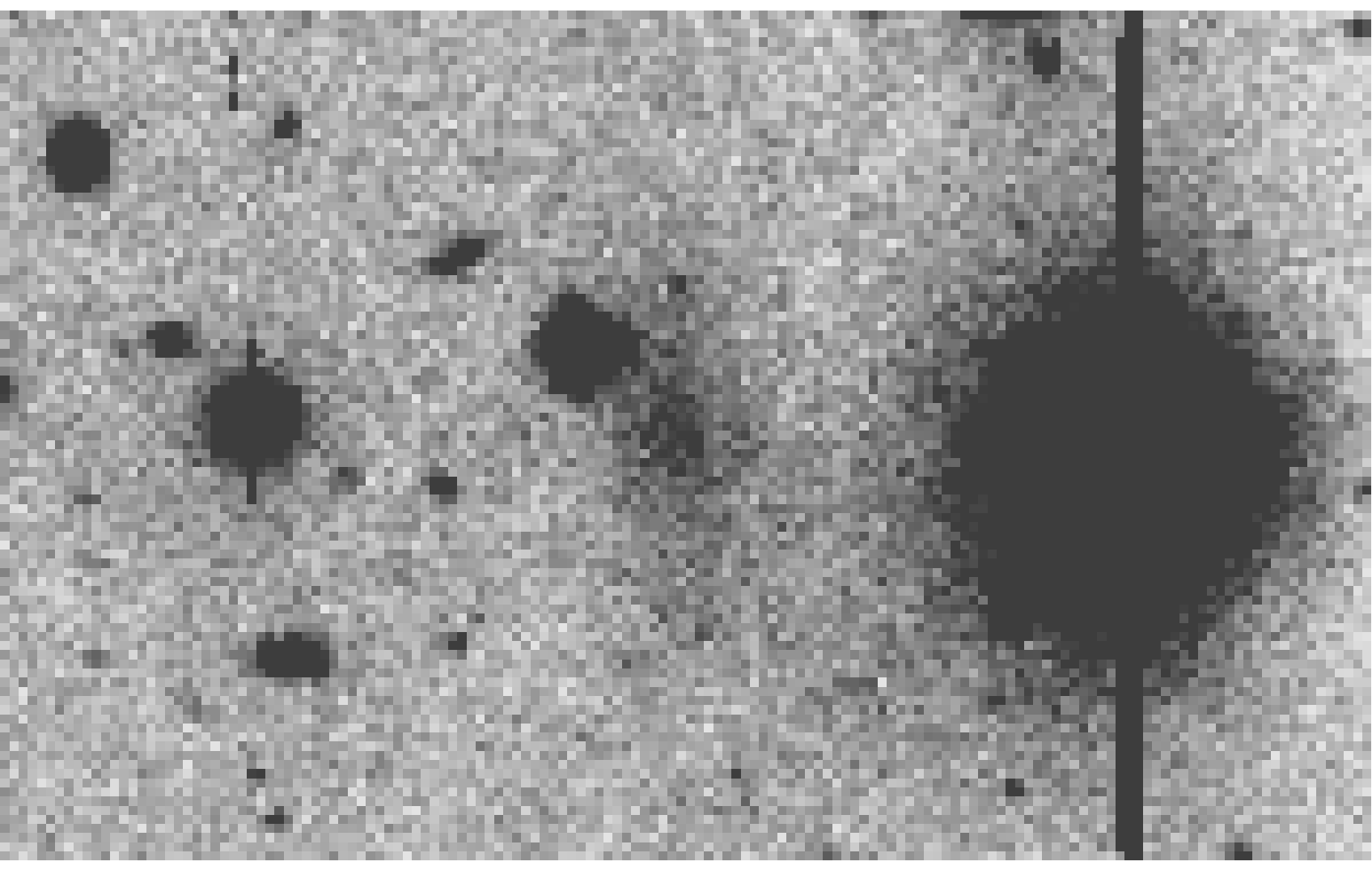,width=5.7cm}\hspace{-5.7cm}
  \parbox[t]{5.7cm}{\vspace{-0.5cm}\hspace{0.6cm}\Large \bf IM4\_2\_LSB1}\hspace{0.1cm}
  \epsfig{figure=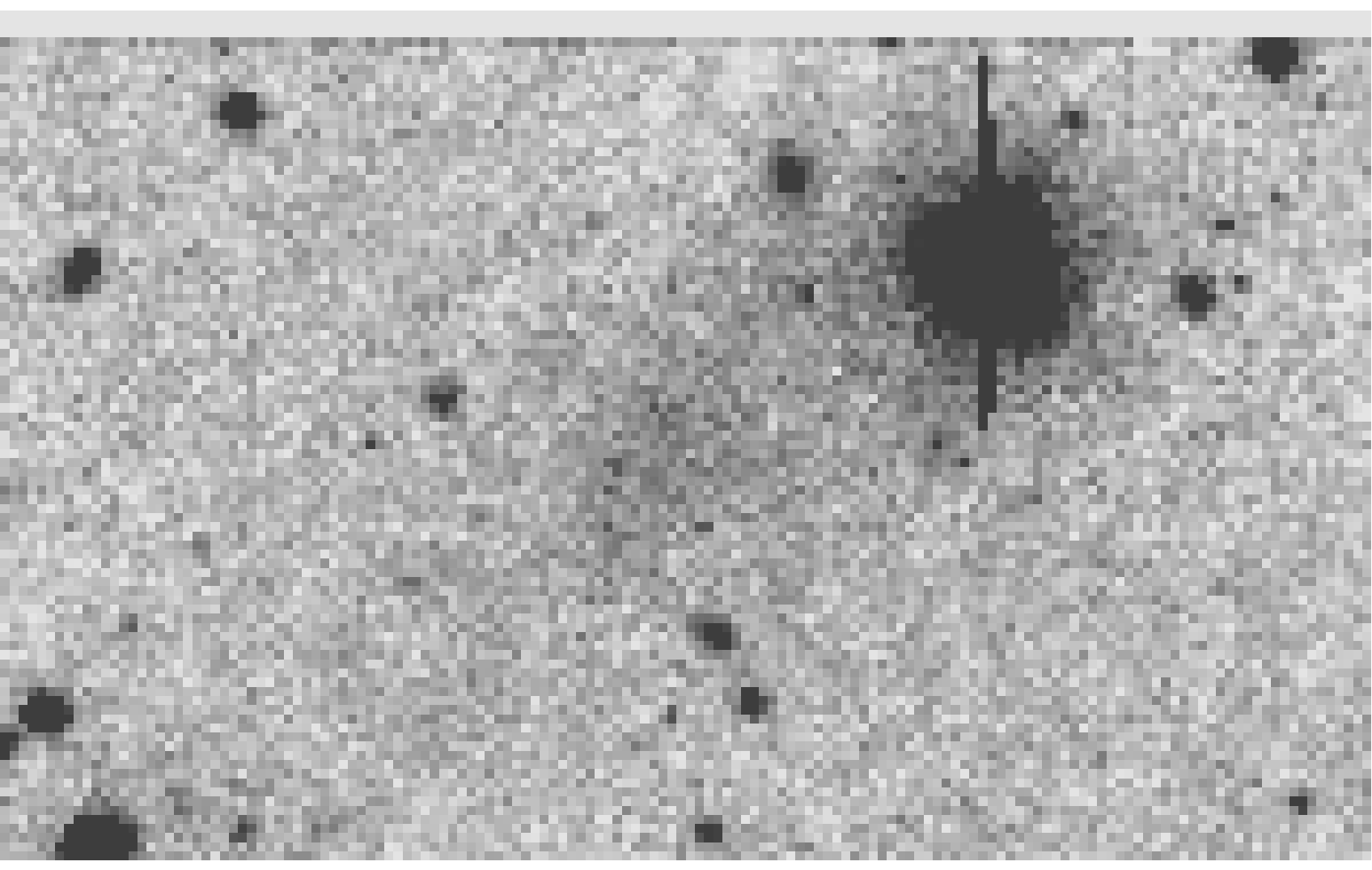,width=5.7cm}\hspace{-5.7cm}
  \parbox[t]{5.7cm}{\vspace{-0.5cm}\hspace{0.6cm}\Large \bf IM7\_2\_LSB1}\hspace{0.1cm}
  \epsfig{figure=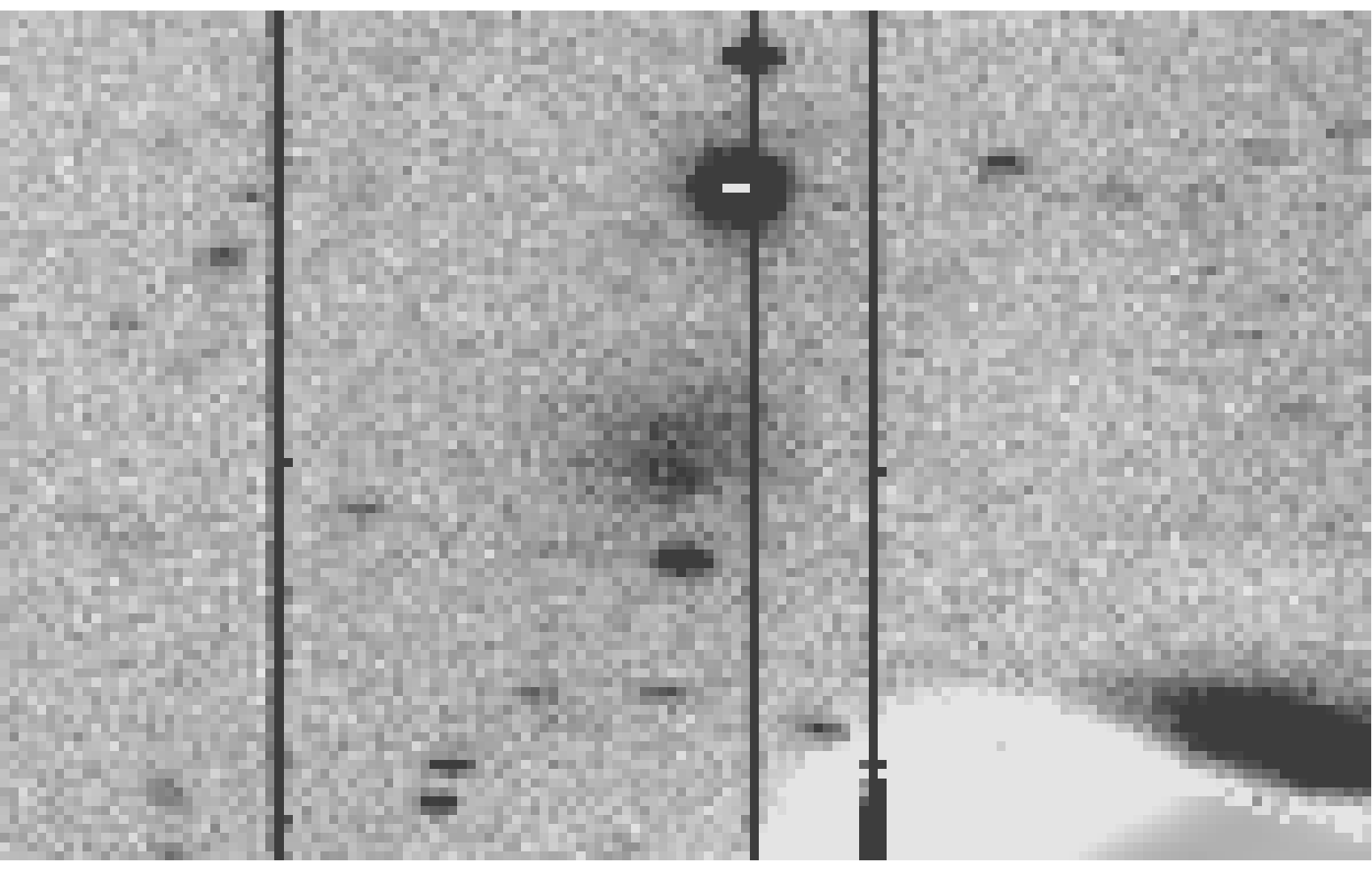,width=5.7cm}\hspace{-5.7cm}
  \parbox[t]{5.7cm}{\vspace{-0.5cm}\hspace{0.6cm}\Large \bf IM1\_6\_LSB2}

  \epsfig{figure=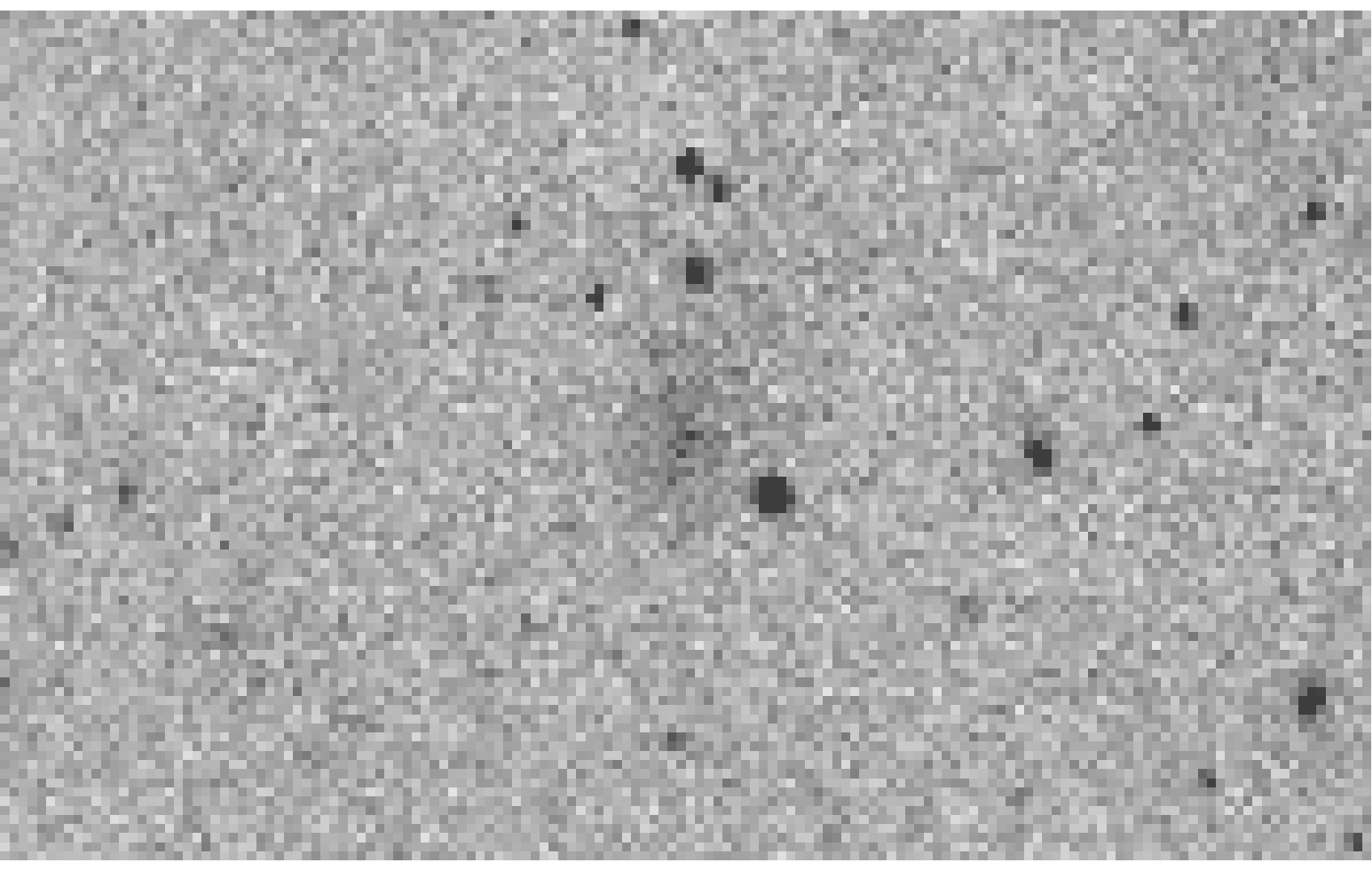,width=5.7cm}\hspace{-5.7cm}
  \parbox[t]{5.7cm}{\vspace{-0.5cm}\hspace{0.6cm}\Large \bf IM1\_2\_5\_LSB1}\hspace{0.1cm}
  \epsfig{figure=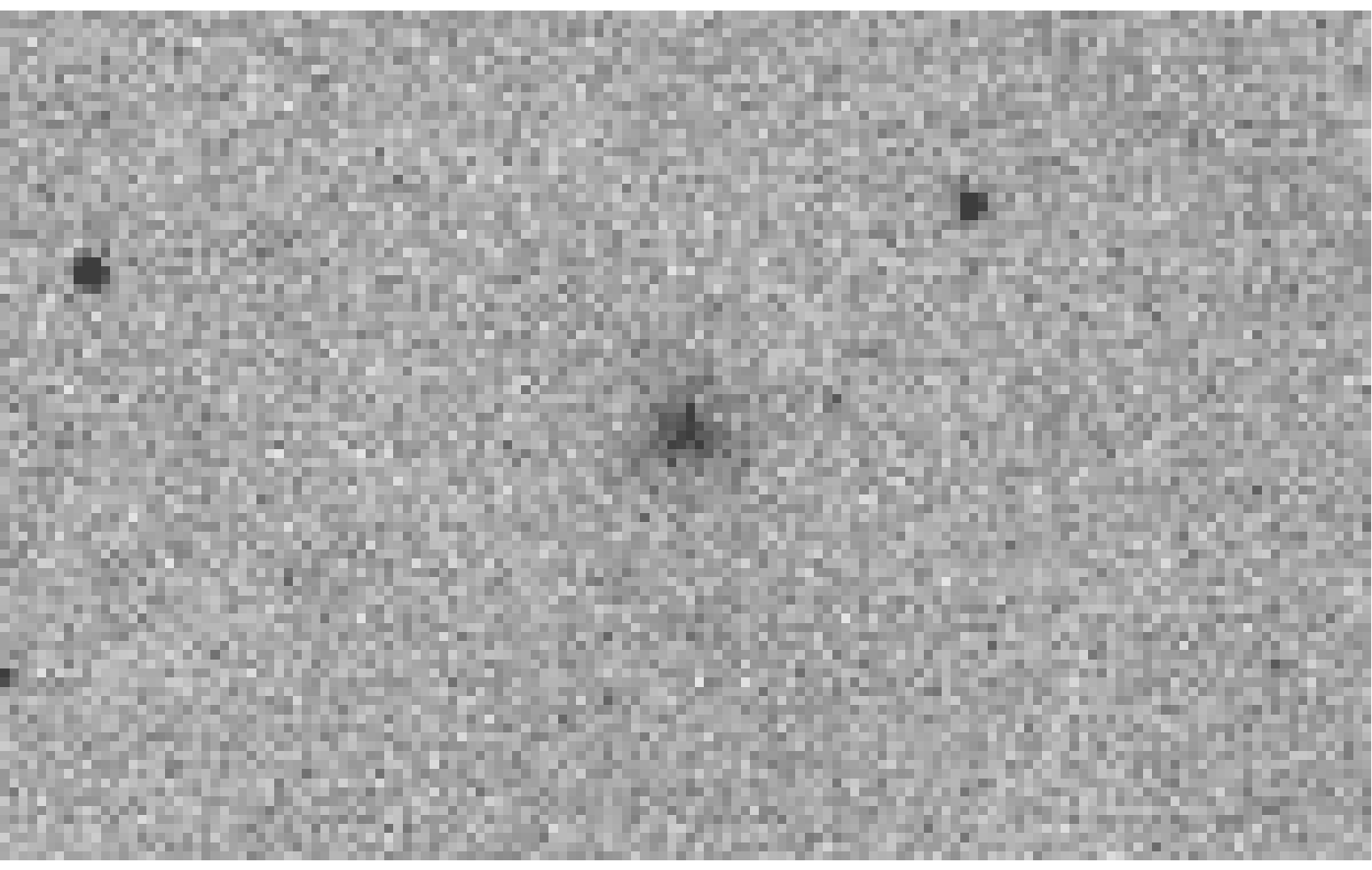,width=5.7cm}\hspace{-5.7cm}
  \parbox[t]{5.7cm}{\vspace{-0.5cm}\hspace{0.6cm}\Large \bf IM1\_2\_2\_LSB1}\hspace{0.1cm}
  \epsfig{figure=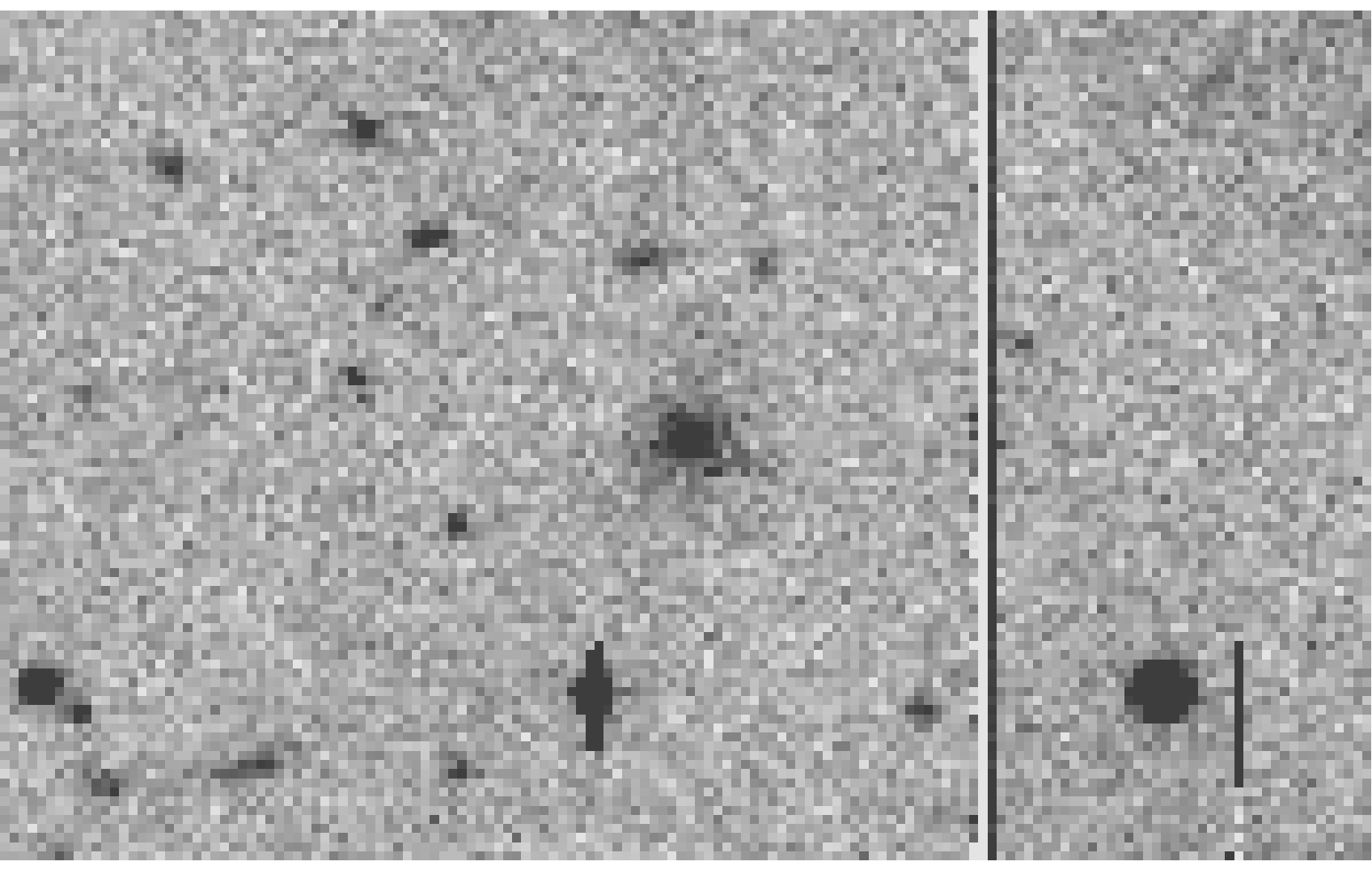,width=5.7cm}\hspace{-5.7cm}
  \parbox[t]{5.7cm}{\vspace{-0.5cm}\hspace{0.6cm}\Large \bf IM7\_6\_LSB1}

  \epsfig{figure=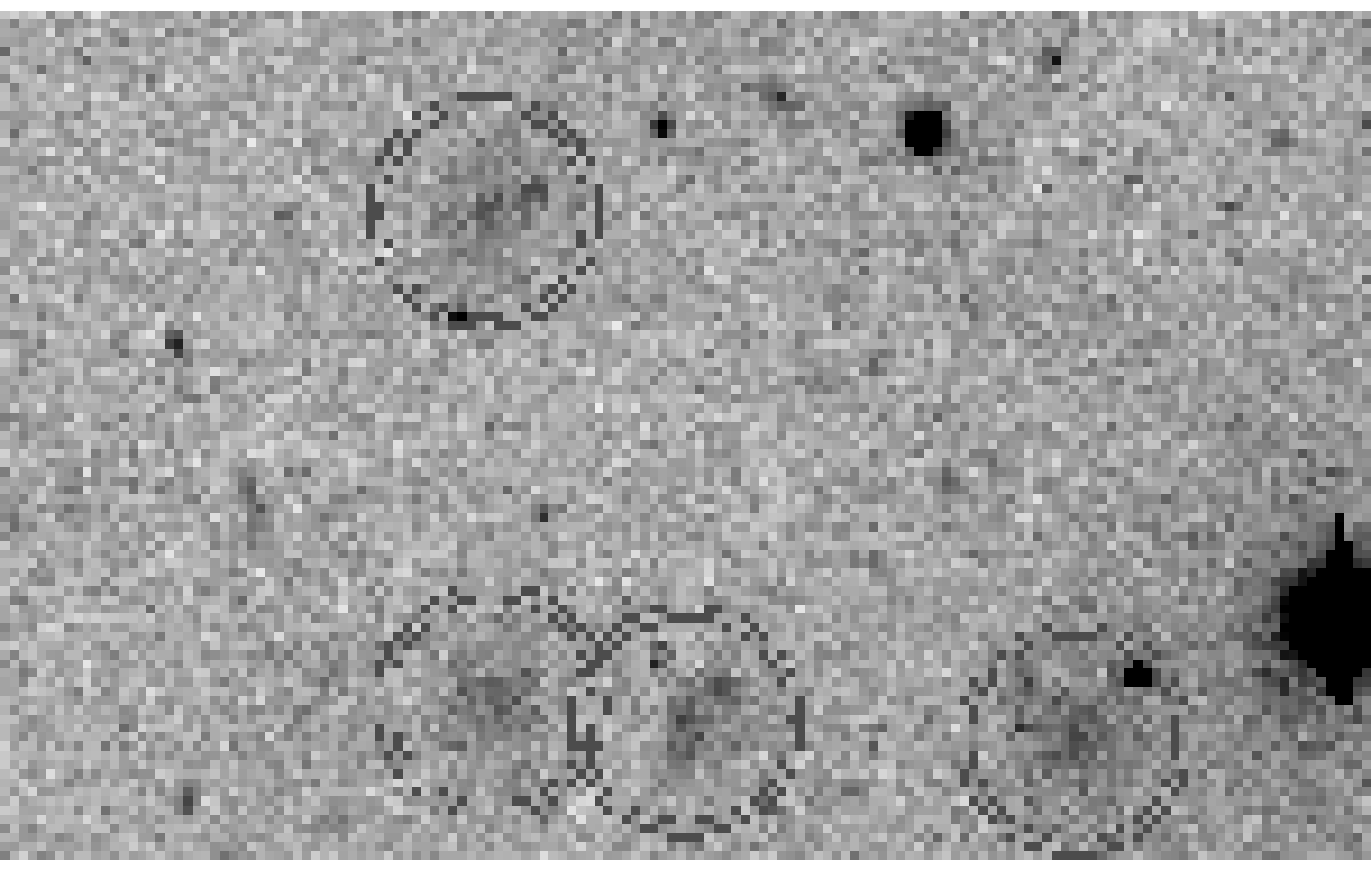,width=5.7cm}\hspace{-5.7cm}
  \parbox[t]{5.7cm}{\vspace{-2.0cm}\hspace{0.3cm}\Large \bf Simulated LG dEs I}\hspace{0.1cm}
  \epsfig{figure=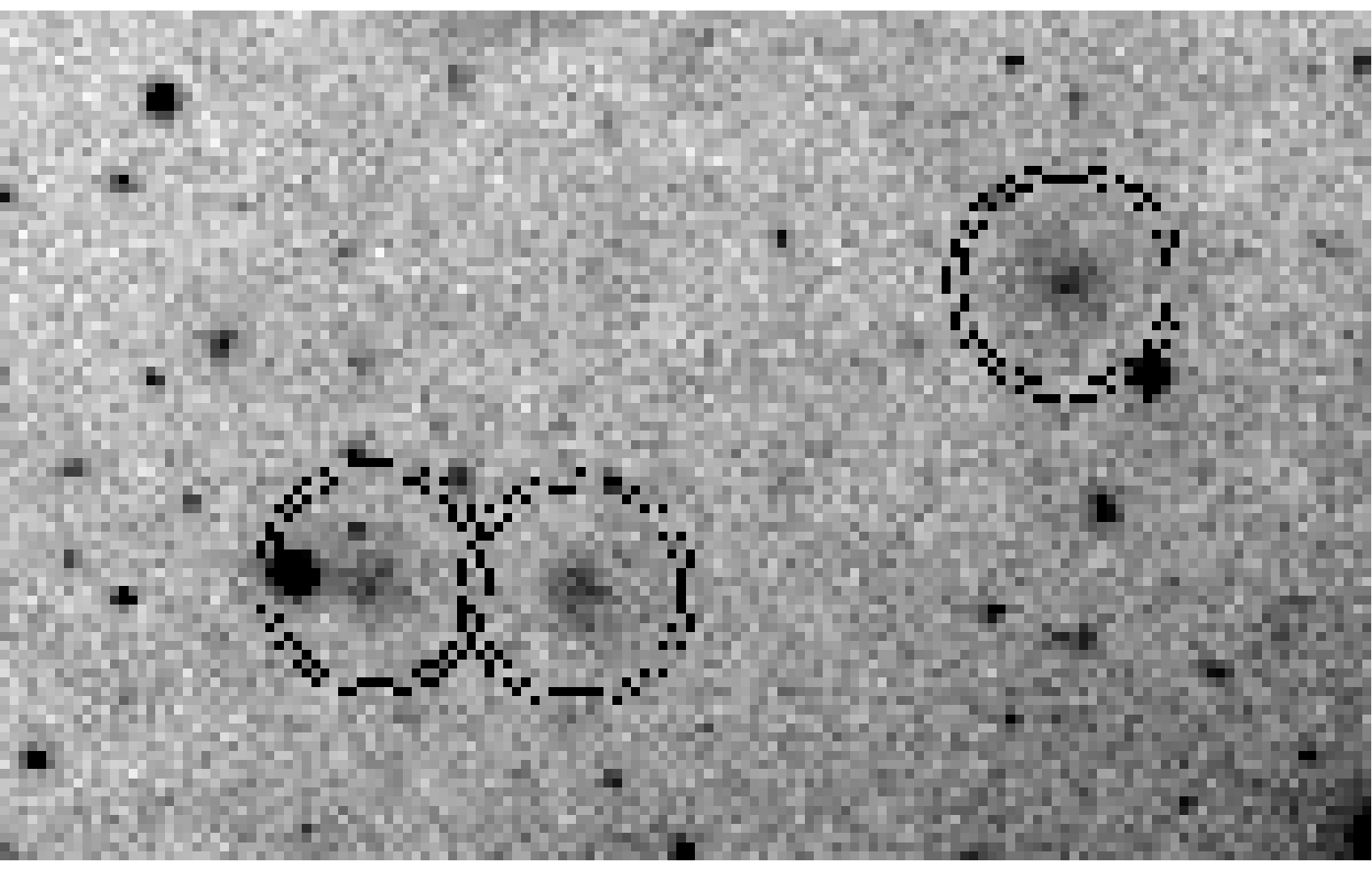,width=5.7cm}\hspace{-5.7cm}
  \parbox[t]{5.7cm}{\vspace{-0.7cm}\hspace{0.3cm}\Large \bf Simulated LG dEs II}\hspace{0.1cm}
  \epsfig{figure=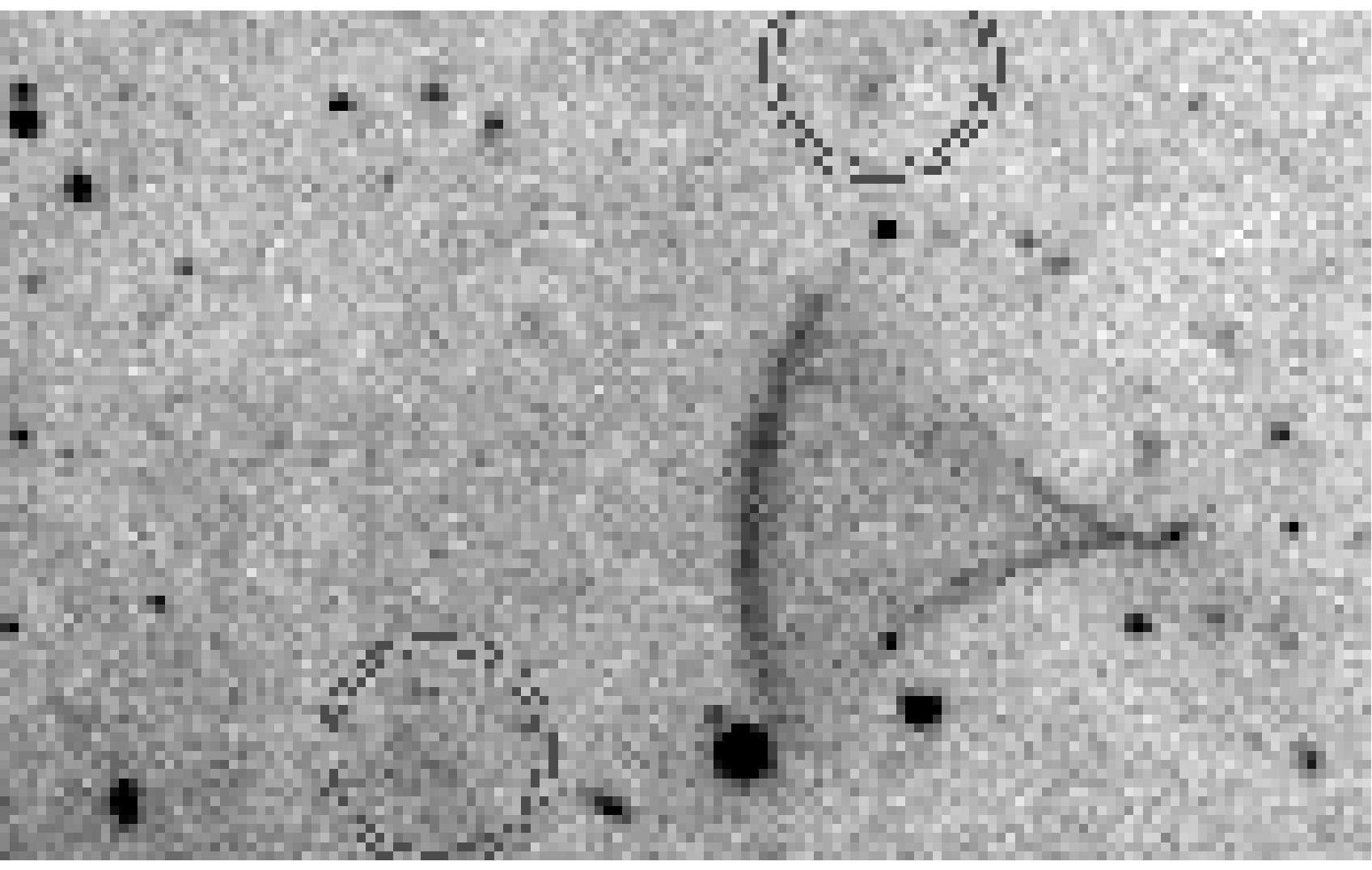,width=5.7cm}\hspace{-5.7cm}
  \parbox[t]{5.7cm}{\vspace{-2.3cm}\hspace{0.5cm}\Large \bf Spurious detections}
 \caption{{\bf Top two rows:} example thumbnails for six of the 12 new Fornax dE candidates detected from our IMACS photometry (Sect.~\ref{search}), see also Table~\ref{sbfresults}. Thumbnails are 82 $\times$ 72$''$ large (7.5 $\times$ 6.6 kpc at the Fornax cluster distance).  They are ordered from left to right and top to bottom by decreasing galaxy luminosity. 
{\bf Bottom row}: the left thumbnail shows four simulated dE galaxies in the magnitude range $-10.8>M_{\rm V}>-11$ mag that have sizes corresponding to the LG dEs (simulated seeing FWHM was 0.9$''$). The middle thumbnail shows three simulated dEs in the same absolute magnitude range, but with only 60\% the size of LG dEs. They are still clearly resolved. The right thumbnail shows two SExtractor detections that were not accepted as new dE candidates, one since it was part of a ghost image, another one because it was a blend of several very small sources. }  
\label{newcands}
\end{center}
\end{figure*}

\section{Search for new members}
\label{search}
The WFCCD data from paper II had a resolution limit which is close to
the intrinsic size of LG dEs for $M_{\rm V}>-12$ mag (see also
Fig.~\ref{mumag}).  The IMACS data now allow us to check how many dE
candidates may have been overlooked in the WFCCD data because of this
restriction.

To do so, we first added several hundred images of artificial dEs of
colour $(V-I)=1.0$ in the range $-12.5<M_{\rm V}<-10$ mag on top of the
IMACS $I$-band images. For simulating them we assumed exponential
surface brightness profiles. Their central surface brightnesses were
adopted to randomly scatter $\pm$ 1.5 mag around the LG
magnitude-surface brightness relation (Grebel et al.~\cite{Grebel03},
see also Fig.~\ref{mumag}). This scatter corresponds to the 2$\sigma$
width of the magnitude-surface brightness relation for the Fornax dE
candidates found in paper II. We then let SExtractor run on those
images to recover the simulated dEs. To achieve efficient detection of
those artifical galaxies and minimize the contamination by more
compact sources, we demanded as detection threshold 400 connected
pixels above 0.7$\sigma$ of the sky noise. Only for the central
pointing with lower integration time we demanded 200 connected pixels.
The detection completeness ranged between 88\% for the brightest dEs
with $-12.5<M_{\rm V}<-12$ mag to 74\% for the faintest dEs with
$-10.5<M_{\rm V}<-10$ mag, with an average of 80\%.

We then examined the SExtractor parameter space covered by those
artificial dEs in terms of isophotal magnitude, surface brightness,
Kron-radius, image area and FWHM. To find new dE candidates, we run
SExtractor on all IMACS images, applying the same detection parameters
as for the IMACS images with added artificial galaxies. We then
restricted the SExtractor output catalog to the parameter space
recovered by the program for the artifical dEs. This selection was
very efficient, resulting in only very few candidates per chip,
sometimes none. We then compared those detected sources
morphologically with the simulated dE. We rejected obvious backgound
objects like spirals, unresolved sources in the region of strong image
distortion, fringing and reflection artefacts, and also groups of
separate neighbouring sources that had not been correctly deblended by
SExctractor (see Fig.~\ref{newcands} for examples). We also
double-checked the image morphology in the shorter exposure $V$-band
images. Obvious V-band dropouts -- hence high z objects -- were
rejected, although those were very few galaxies, all restricted to
compact relatively high-surface brightness objects. In addition to the
automated SExtractor search, we finally inspected the images visually
to search for sources morphologically representing the simulated LG dE
analoga.

\begin{figure}
\begin{center}
  \epsfig{figure=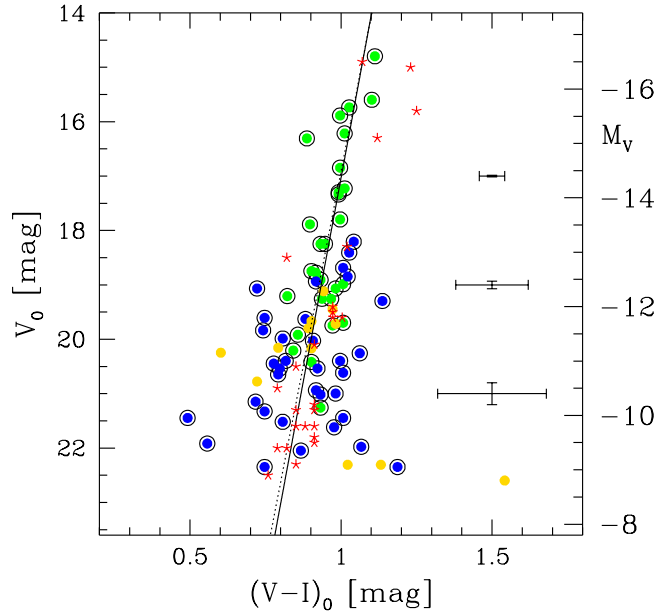,width=8.6cm}
       \caption{Colour-magnitude diagram (CMD) of the sources from Fig.~\ref{mumag_imacs}. Typical error bars are indicated. The magnitudes and colours were redenning corrected using Schlegel et al.~(\cite{Schleg98}). The $V$ magnitude is from the IMACS data, while  $(V-I)$ is adopted as the mean of the WFCCD and IMACS values. Adopting this mean reduces the colour scatter by about 35\% as compared to the single IMACS colour: the colour scatter for the IMACS values is 0.15 mag, while that of the mean (and also the WFCCD colour) is about 0.11 mag. The solid line is a fit to the data points applying a 3$\sigma$ clipping.  The dotted line -- which is practically on top of the solid line -- indicates the colour-magnitude relation derived from the WFCCD photometry of all dE candidates (Hilker et al.~\cite{Hilker03}, paper II). The (red) Local Group data points for $M_{\rm V}<-13$ mag are proper $(V-I)$ measurements from Mateo et al.~\cite{Mateo98}, while for $M_{\rm V}>-13$ mag the $(V-I)$ values are estimated from their [Fe/H] values (Grebel et al.~\cite{Grebel03}). For this transformation we applied equation (4) from Kissler-Patig et al.~(\cite{Kissle98}), and adopted an additional zero point shift of -0.055 mag. This shift was the difference between the transformed and measured $(V-I)$ values for the five Local Group dEs with $M_{\rm V}<-13$ mag that do have a measured $(V-I)$.}
\label{cmd}
\end{center}
\end{figure}

This search resulted in the detection of 12 dE candidates in the range
$-12.3<M_{\rm V}<-8.8$ mag that had not been detected in the WFCCD data in
paper II. Example images are shown in Fig.~\ref{newcands}. Eight of
the 12 new candidates were detected by SExtractor, four additional
ones by visual inspection only (see Table~\ref{sbfresults}). This
fraction of visual detections is consistent with SExtractor's
incompleteness as quoted above.  The locations of all new dE
candidates are indicated in the Fornax map (Fig.~\ref{map}). Their
photometric parameters are listed in Table~\ref{sbfresults} and shown
graphically in Figs.~\ref{mumag_imacs}~and~\ref{cmd}. Four of these 12
dE candidates are located outside the image borders of the WFCCD data,
all of which are included in the FCC as likely cluster members (see
Table~\ref{sbfresults}). There is furthermore one new dE candidate
listed as probable background galaxy in the FCC (see
Table~\ref{sbfresults}).

How many of the new dE candidates can be confirmed via SBF? We
detected an SBF signal for only two galaxies, sources IM4\_2\_LSB1 and
IM1\_2\_2\_LSB1. The first galaxy has the highest total luminosity of
the new dE candidates, while the second galaxy has the highest surface
brightness. We have applied the SBF reduction procedure from paper III
to these galaxies. The resulting SBF parameters are listed in
Table~\ref{sbfnew}. The $S/N$ of the SBF signal was 5.0 for
IM4\_2\_LSB1 and 1.9 for IM1\_2\_2\_LSB1. Their SBF distances are
marginally consistent with the Fornax cluster distance of 31.39 mag
(Freedman et al.~\cite{Freedm01}), albeit at the low limit. We accept
both galaxies as SBF confirmed Fornax cluster member, but assign
IM1\_2\_2\_LSB1 an intermediate membership flag=1.3 due to its low
$S/N$ in the SBF measurement (see Table~\ref{sbfresults}).

None of the other new candidates had a measurable SBF signal. It must
therefore be checked whether among those may be any bona-fide SBF
background galaxy. From Fig.~\ref{mumag_imacs} it is clear that apart
from the two candidates with SBF signal, six more have central surface
brightnesses $<24.5$ mag/arcsec$^2$, which was the approximate SBF
detection limit for the entire sample of dE candidates from paper II.
Three out of those six are detected in the {\it central} dithered
IMACS pointing which only had 1800 seconds total exposure time. Two of
the galaxies, namely IM1\_2\_6\_LSB1 and IM1\_1\_7\_LSB1, are indeed
detected in only one of the two dithered exposures of 900 seconds
integration time. This substantial reduction of integration time by a
factor of 2-4 compared to the other fields results in a brighter
limiting magnitude for SBF measurement by 1-2 mag (see paper I), hence
explaining the absence of an SBF signal. Two further galaxies
IM1\_6\_LSB1 and IM1\_6\_LSB2 had very large PSF FWHM $\simeq$
1.6$''$, being located in a part of the field of view with strong
image distortion (see Fig.~\ref{newcands}). This leaves one single
object, namely IM7\_6\_LSB1, as a galaxy with reasonably good seeing
and long $I$-band integration time. Fig.~\ref{newcands} shows that
this galaxy is very compact, in fact yielding too few independent data
points for SBF sampling. Furthermore, this galaxy has the bluest
colour of all new dE candidates (see next Section and Fig.~\ref{cmd}).
It may therefore be a background blue compact dwarf or unresolved
background spiral. We have assigned this galaxy an intermediate
cluster membership flag of 2.3 (see Table~\ref{sbfresults}).

\section{The revised photometric properties of Fornax early-type dwarf galaxies}
\label{revision}
In this section we discuss the photometric properties of the early
type Fornax dwarf galaxy population based on the IMACS photometry. We
include in our analysis those galaxies from Table~\ref{sbfresults}
with membership flag $<2.5$. They are the ones directly classified as
cluster members from SBF (Sect.~\ref{SBF}), plus those that are
probable cluster members from our revised morphological assessment
(Sect.~\ref{morph}), plus those that were newly discovered in the
IMACS data (Sect.~\ref{search}).

Total galaxy magnitudes were derived with the IMACS data as for the
WFCCD data by curve-of-growth analyses using the IRAF task ELLIPSE.
Along each isophote fitted by ELLIPSE, a 3$\sigma$ clipping algorithm
was applied to reject contaminating sources. The sky level was
adjusted individually for every galaxy in the course of the
curve-of-growth analysis. Total magnitudes were derived by summing up
the fitted intensities up to a cutoff radius determined by the
curve-of-growth analysis. Central surface brightnesses were determined
by fitting an exponential function to the surface brightness profile,
excluding the nuclear regions of dE,Ns. 

Colours from the IMACS data are derived as the difference between $V$
and $I$ magnitudes within an aperture of 4$''$ radius, exactly like
for the WFCCD data. We adopt as the final galaxy colour the mean of
the WFCCD and IMACS values, given that the scatter in the
colour-magnitude plane reduces by 35\% when doing so. Zero-point
colour offsets between both photometry sets are negligible: the mean
colour difference between IMACS and WFCCD data is $-0.006 \pm 0.026$
mag. 

It is
  worth detailing the procedure applied to obtain a realistic colour
  error for each galay. The first step was to estimate a global
  uncertainty of the sky background determination for the IMACS data.
  The corresponding colour error then is robust in a relative sense
  such that galaxies with fainter surface brightness have
  correspondingly larger colour errors.  For the global sky background
  uncertainty in the IMACS data we adopted the RMS scatter between two
  sets of estimates: first, the background obtained ``manually'' from
  the curve-of-growth analysis.  Second, the sky background obtained
  from subtracting a SExtractor sky map off the galaxy image, which
  itself had previously been cleaned of all objects using a SExtractor
  object map. 

To obtain a realistic colour error also in an absolute
  sense, we re-scaled the estimated global sky background uncertainty
  such that the resulting average colour error equals the RMS scatter
  between the IMACS and WFCCD colours. The colour errors derived in
  this way were on average around 0.10 mag, with a broad range between
  0.02 mag for the highest surface brightness galaxies and almost 0.40
  mag for the faintest ones (see Table~\ref{sbfresults}).

For $\mu_{\rm V,0}$ and $M_{V,\rm 0}$, we use only the IMACS values,
given that the scatter of the mean values in the magnitude-surface
brightness plane is marginally larger than the scatter of the IMACS
values alone. Errors in $\mu_{\rm V,0}$ and $M_{V,\rm 0}$ were
  derived from the uncertainty in the sky background determination in
  the IMACS data, as outlined in the previous paragraph. The resulting
  errors are given in Table~\ref{sbfresults}. We finally note that
the IMACS $\mu_{\rm V,0}$ and $M_{V,\rm 0}$ values are on average
about 0.1 mag fainter than the WFCCD values. This is because the
higher resolution IMACS data allowed a better masking of contaminating
point sources close to the galaxy centres, especially for the lower
surface brightness galaxies.

\subsection{Magnitude-surface brightness relation}
\label{magmu}
In Fig.~\ref{mumag_imacs} we show the IMACS magnitude-surface
brightness plot, with typical error bars indicated. 
Data for LG dEs are shown for comparison.
We fit the following relation between
central surface brightness $\mu_{\rm V,0}$ and absolute magnitude
$M_{V,\rm 0}$, assuming $(m-M)=31.39$ mag and applying a 3$\sigma$
clipping algorithm:\footnote{The fit errors are derived from random
  resampling of the data points within their measured scatter. Given
  that the scatter of values in Fig.~\ref{mumag_imacs} is much larger
  than the measurement uncertainty in both $V_{\rm 0}$ and $\mu_{\rm
    V,0}$, we do not error weight the data points in the fit.}
\begin{equation}
\mu_{\rm V,0}=32.32 (\pm 1.12) + 0.681 (\pm 0.040) \times M_{V,\rm 0}
\label{mumagrel}
\end{equation}

Assuming the same slope, the relation defined by the LG dEs has an
offset of 0.75 $\pm$ 0.15 mag towards brighter $\mu_{\rm V,0}$. In
other words, the Fornax dEs in our sample are about 40\% larger than
the LG dEs at equal central surface brightnesses.  There have been
indications for such a size difference also for Virgo cluster dwarfs:
Caldwell \& Armandroff~(\cite{Caldwe00}) and
Caldwell~(\cite{Caldwe05}) report on the discovery of a large
population of very low surface brightness dEs in the Virgo cluster
whose sizes extend to significantly larger values than known for LG
dEs.

An offset in mean size may be explained within a scenario where tidal
forces disrupt the smallest and least massive dEs more effectively in
denser environment like the Fornax or Virgo cluster than in the LG
(e.g. Hilker et al.~\cite{Hilker99c}). However, Fig.~\ref{deltamu}
suggests that the Fornax cluster also hosts an overabundance of larger
dEs, in addition to an underabundance of smaller ones. Indeed, the
surface brightness distribution of LG and Fornax totally agree with
each other once a simple offset to the LG values is applied, see
Fig.~\ref{deltamu}. One may therefore speculate that tidal heating
(e.g. Valluri~\cite{Vallur93}, Das \& Jog~\cite{Das95}) could be
enhancing the internal energy of dEs in the Fornax cluster more than
for LG dEs. This is conceivable since tidal heating effects are
expected to be more pronounced in denser environments
(Valluri~\cite{Vallur93}), as is the Fornax cluster in comparison with
the LG. It remains to be clarified whether the vastly dark matter
dominated faint dEs could indeed be sufficiently affected by the
cluster tidal field.

Another mechanism that may in principle cause environmental
differences among galaxy populations is re-ionization (e.g. Dekel \&
Woo~\cite{Dekel03}, Moore et al.~\cite{Moore06}). It is reasonable to
assume that those galaxy halos that collapsed in the densest regions
of the universe collapsed very early, and consequently had more time
to form stars before re-ionization than halos in less dense regions of
the universe. This may have led to less centrally concentrated stellar
halos in galaxies located in high density environments like Fornax.
However, Grebel \& Gallagher~(\cite{Grebel04}) line out that the star
formation histories of most LG dwarfs were probably not decisively
influenced by re-ionization. Along these lines, also Ricotti et
al.~(\cite{Ricott02}) argue that the star formation histories of low
mass dark matter haloes are almost independent of the external
radiation field and mostly influenced by radiative feedback.

A general concern about the comparison between the LG and other
environments is that the LG sample may be highly incomplete. In that
context it is interesting to note that the latest discoveries of new
faint LG members with $M_{\rm V}\simeq-$ 9 mag (e.g. Armandroff et al.
\cite{Armand99}, Whiting et al. \cite{Whitin99}, Zucker et
al.~\cite{Zucker04}) are restricted to rather small scale sizes. This
may indicate that the LG dE sample is more incomplete at smaller than
at larger galaxy sizes. The fact that {\it large scale} stellar
overdensities like the Sagittarius and Monoceros stream {\it are}
quite readily detected in all-sky surveys (Ibata et
al.~\cite{Ibata01}, Majewski et al.~\cite{Majews04}, Pe\~narrubia et
al.~\cite{Penarr05}), seems to support this impression.

It is clear that more work still needs to be done regarding this
subject: both regarding a comparison of structural parameters of LG
dwarfs with their counterparts in nearby clusters, and, the
theoretical framework of dwarf galaxy formation.

\subsection{Colour-magnitude relation}
\label{cmdsec}
In Fig.~\ref{cmd} we show the colour-magnitude diagram of the same
sources as in the magnitude-surface brightness plot in
Fig.~\ref{mumag_imacs}, with typical error bars indicated. 
Data for LG dEs are also shown for comparison. We fit
the following relation between colour $(V-I)_{\rm 0}$ and absolute
magnitude $M_{V,\rm 0}$, applying a 3$\sigma$ clipping
algorithm\footnote{For this fitting we weight each data point by its
  colour error, given that the scatter in the CMD is consistent with
  being entirely created by colour measurement errors.}
\begin{equation}
(V-I)_{\rm 0}=0.52 (\pm 0.07) - 0.033 (\pm 0.004) \times M_{V,\rm 0}
\label{vimagrel}
\end{equation}

The colours of Fornax dEs are in reasonable agreement with the values
for Local Group dEs except for the brightest few galaxies, indicating
overall similar self-enrichment histories for both populations. There
is a marginal shift of 0.032 $\pm$ 0.014 mag towards redder colours
for the LG data. However, it must be noted that the LG dE colours for
$M_{\rm V}>-13$ mag are no direct measurements but rather estimated from
their [Fe/H] values (see caption of Fig~\ref{cmd}). When restricting
the consideration to $M_{\rm V}<-13$ mag the colour difference is
substantially larger with 0.14 $\pm$ 0.05 mag. The nominal slope of
the colour-magnitude correlation for the LG dEs is $-0.047 \pm 0.005$,
about 40\% higher than for the Fornax dEs at 2.2$\sigma$ significance.
We note that these differences are consistent with a higher
self-enrichment efficiency among LG dEs than in Fornax dEs --
especially in the brighter luminosity regime. It is an interesting
speculation that the more compact light distribution of LG dEs (see
previous sub-section) may be part of the reason for such a stronger
self-enrichment, provided that light traces mass in a similar manner
in both environments. Both a stronger gravitational potential and a
higher gas density may have supported self-enrichment in Local Group
dEs more than in Fornax dEs. The smaller sizes of Local Group dEs and
the (still preliminary) colour differences between LG and Fornax dEs
could therefore be two sides of the same medal.
\begin{figure}
\begin{center}
  \epsfig{figure=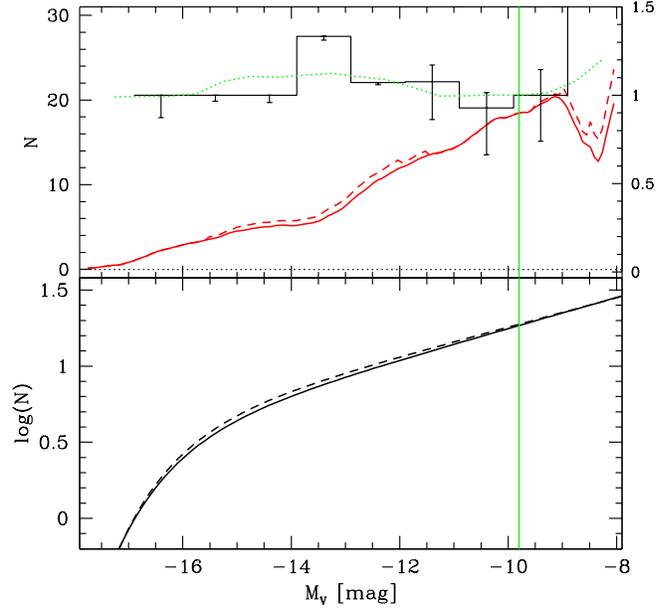,width=8.6cm}
       \caption{This figure shows how the Fornax cluster galaxy
         luminosity function from Hilker et al.~(\cite{Hilker03},
         paper II) is re-assessed in the present paper. The vertical
         (green) line indicates the 50\% detection limit from paper
         II, which is also the faint limit for fitting the faint end
         slope $\alpha$. {\bf Top panel:} The solid (red) line
         indicates a smoothed representation of the Fornax GLF from
         paper II. The solid histogram gives the fraction of correct
         cluster member candidate identifications determined in
         Sects.~\ref{SBF} to ~\ref{morph}, augmented by the inclusion
         of the new candidate cluster members from Sect.~\ref{search}.
         Note that this inclusion of members that were undetected in
         paper II makes the fraction larger than 1 in some bins. The
         (green) dotted line is a smoothed representation of the
         histogram. The dashed red line indicates the corrected Fornax
         GLF, i.e. the product of the paper II LF with the dotted
         line. See Sect.~\ref{GLF} for more details. {\bf Bottom
           panel:} Solid line: 1-component fit to the Fornax-GLF from
         paper II. Dashed line: 1-component fit to the {\it corrected}
         Fornax GLF.  The respective faint end slopes are
         $\alpha=-1.10$ for the paper II LF and $\alpha=-1.09$ for the
         corrected LF. }
\label{lf_fdf}
\end{center}
\end{figure}
\subsection{Galaxy luminosity function}
\label{GLF}
In Fig.~\ref{lf_fdf} we show the Fornax dE GLF, as derived from
scaling the paper II GLF with the fraction of galaxies that are
confirmed as cluster members in this paper. Note that the inclusion of
the new candidates from Sect.~\ref{search} raises this fraction above
unity in some magnitude bins.

The denominator of this fraction is the number of galaxies per
magnitude bin from paper II that enter in the calculation of the paper
II LF. Note that these are only those that lie within the 2$\sigma$
limits of both the colour and surface-brightness magnitude relation
defined by the entire sample. This restriction had been applied in
paper II to reduce the effect of contaminators to the sample, given
the limited morphological selection potential of the WFCCD data
especially in the parameter space defined by LG dEs. The IMACS data
largely removes this restriction by virtue of its improved spatial
resolution. For the nominator we hence adopt the number of confirmed
cluster member candidates -- i.e. those with flag $<$ 2.5 in
Table~\ref{sbfresults} -- {\it without} applying any cut. We do
however exclude those five new dE candidates that were outside the
WFCCD FOV.

Applying a one component Schechter function fit to the resulting dwarf
galaxy luminosity function down to $M_{\rm V}=-9.8$ mag (the 50\% WFCCD
completeness limit), the faint end slope is $\alpha=-1.10 \pm 0.10$
both for the paper II GLF and for the corrected GLF (see
Fig.~\ref{lf_fdf}). Note that the paper II GLF and hence also the
corrected GLF is incompleteness corrected. The nominal slope
difference is below 0.01. Also the inclusion of the five new dE
candidates outside the WFCCD FOV does not alter the slope by more than
0.01. Restricting the fit to the very faint end $M_{\rm V}>-13.5$ mag (the
onset of the dSph regime, see Grebel et al.~\cite{Grebel03}), the
values are $\alpha=-1.06$ for the paper II GLF and $\alpha=-1.04$ for
the corrected GLF. The corrected slope is only marginally shallower
than the value from paper II, much smaller than the statistical
uncertainty of $\pm$ 0.10.

It is comforting that the faint end slope of the Fornax dwarf GLF
remains in the range $-1.1$ to $-1.0$ when going from the early
studies of Ferguson \& Sandage~(\cite{Fergus88}) and
Ferguson~(\cite{Fergus89}) to the present paper which samples the GLF
to about 3 magnitudes fainter. We are therefore confident that this
value is robust and not much biased by systematic effects.
Morphological cluster membership assignment in Fornax apparently is
very reliable, provided that the image resolution is sufficient.

It is well known that such a shallow faint end slope is in sharp
contradiction to the much steeper value predicted for the mass
function of $\Lambda$CDM halos (e.g.Kauffman et al.  \cite{Kauffm00},
Moore et al.~\cite{Moore99}). Possible reasons to explain that
discrepancy include the accretion scenario (e.g. Hilker et
al.~\cite{Hilker99c}, C\^{o}t\'{e} et al.~\cite{Cote98}), where dwarf
galaxies that fall into the cluster centres are tidally disrupted,
hence contributing to form the extended cD halos of the most massive
cluster galaxies like NGC 1399 in Fornax. The presence of
ultra-compact dwarf galaxies (UCDs) in the central Fornax cluster
(Hilker et al.~\cite{Hilker99b}, Drinkwater et al.~\cite{Drinkw03})
may be a signpost of these tidal interactions, given that some -- but
possibly not most -- of them may be tidally stripped dE,Ns (Bekki et
al.~\cite{Bekki03}, Mieske et al.~\cite{Mieske06a}). Another
possibility is that the low surface brightness dEs that we see
nowadays originate from tidally stripped originally much more massive
dark matter halos (Stoehr et al.~\cite{Stoehr02}, Kravtsov et
al.~\cite{Kravts04}), while the lower mass halos have not been able to
maintain condensed gas to from stars. However, Kazantzidis et
al.~(\cite{Kazant04}) argue that observed velocity dispersion profiles
of Local Group dE exclude such very massive progenitors, indicating
that tidal stripping can only be a part of the picture. More work on
the theoretical side is certainly needed to better understand the
mechanisms that create and destroy dark matter dominated stellar
systems at the low mass end (e.g. Hayashi et al.~\cite{Hayash03},
Moore et al.~\cite{Moore06}).
\section{Summary and conclusions}
In this paper we have presented a photometric analysis of the
early-type dwarf galaxy population ($M_{\rm V}>-17$ mag) in the central
Fornax cluster, covering the central square degree and slightly
beyond. The analysis is based on wide field imaging in $V$ and $I$
using the instrument IMACS mounted at the 6.5m Walter Baade Magellan
telescope at Las Campanas Observatory, Chile. The pixel scale of
0.2$''$ and 0.8$''$ median seeing FWHM enabled us to efficiently
resolve LG dE analoga down to $M_{\rm V}\simeq -10$ mag. We used these data
to follow up a previous imaging survey of our group (Hilker et
al.~\cite{Hilker03}, paper II) that had four times larger pixel scale
and seven times smaller light collecting area.

We summarize our main results as follows:
\begin{enumerate}
\item We confirm the cluster member status for 28 candidate dEs from
  paper II in the magnitude range $-16.6<M_{\rm V}<-10.1$ mag by means of
  $I$-band surface brightness fluctuation (SBF) measurements.
\item We re-assess the morphological classification of 51 further
  candidate dEs from paper II in the range $-13.2<M_{\rm V}<-8.6$ mag based
  on the much improved imaging resolution. Of these, 2/3 retain their
  smooth dE like appearence on our images, hence are confirmed as
  probable cluster members. About 1/3 are re-classified as probable
  background galaxies, most of which have sizes close to the
  resolution limit of the data from paper II.
  We cannot confirm the status of background galaxy by means of SBF
  measurement. This is because the galaxies without a measurable SBF
  signal are at about or fainter than the limiting surface brightness
  for SBF detection at the Fornax cluster distance.
\item We find 12 new dE candidates in the range $-12.3<M_{\rm V}<-8.8$ mag
  whose intrinsic sizes are close to the resolution limit of the data
  from paper II. These detections result from a search for low surface
  brightness features trimmed to detect analoga of Local Group dEs
  with $M_{\rm V}>-12.5$ mag. Two of the new candidates can be confirmed via
  SBF measurement.
\item We investigate the surface brightness-magnitude relation for the
  joint sample of confirmed dE candidates from items (1) to (3). We
  derive the following fit:\vspace{0.2cm}\\\vspace{0.2cm}$\mu_{\rm
    V,0}=32.32 + 0.681 \times M_{V,\rm 0}$\\ We find that the Fornax
  dEs are shifted by 0.75 $\pm$ 0.15 mag towards fainter $\mu$ at a
  given luminosity compared to Local Group dEs. That is, Fornax dEs
  are about 40\% larger than Local Group dEs. We briefly discuss
  possible reasons for this difference.
\item We investigate the colour-magnitude relation for the same sample
  of Fornax dEs and find the
  following fit:\vspace{0.2cm}\\\vspace{0.2cm}$(V-I)_{\rm 0}=0.52 - 0.033\times M_{V,\rm 0}$\\
  The derived slope is slightly shallower than estimated for Local
  Group dEs, mainly driven by the fact that Local Group dEs with
  $M_{\rm V}<-13$ mag appear significantly redder than their Fornax
  counterparts of same luminosity. We indicate that in the context of
  self-enrichment this may be expected from the fact that Local Group
  dEs have a more compact stellar body than Fornax dEs.
\item We re-scale the paper II dwarf galaxy luminosity function (GLF)
  in Fornax as a function of magnitude by the fraction of confirmed
  candidate dEs derived in items (1) to (3). A one-component Schechter
  fit of the corrected GLF down to $M_{\rm V}=-9.8$ mag yields a faint end
  slope $\alpha$ differing by less than 0.01 from the value
  $\alpha=-1.10 \pm 0.10$ derived in paper II. When restricting the
  fit to $-13.5<M_{\rm V}<-10$ mag, the slope is $\alpha=-1.04$ for the
  corrected GLF and -1.06 for the paper II GLF. Our results confirm a
  very shallow faint end slope for the Fornax dwarf galaxy luminosity
  function, in agreement with early estimates in the reference study
  of Ferguson \& Sandage~(\cite{Fergus88}). We briefly discuss this
  finding in the context of structure formation models.
\end{enumerate}
We conclude that for nearby clusters such as Fornax, the SBF method is
a powerful tool for extending the limit of cluster membership
determination down to the regime where the faint end slope $\alpha$
dominates the shape of the GLF. Our predictions for the potential of
the SBF method (Mieske et al.~\cite{Mieske03}, paper I) are confirmed
by the results of this paper. We can also conclude that for the Fornax
cluster case, the morphological cluster membership assignment is very
efficient, provided that the image resolution is substantially better
than the expected galaxy size. Future targets for studies like the one
presented here include the Virgo cluster, or other nearby galaxy
groups within about 20 Mpc.
\label{summary}
\begin{acknowledgements}
  We thank the staff at Las Campanas Observatory for their friendly
  and very efficient support during the execution of the imaging runs.
  SM acknowledges support by DFG project HI 855/1 and DAAD Ph.D. grant
  Kennziffer D/01/35298. LI was supported by FONDAP ``Center for
  Astrophysics''. CMdO would like to thank the
  Universit\"atssternwarte M\"unchen and the Max Planck Institut f\"ur
  Extraterretrische Physik for their hospitality. CmdO acknowledges
  support from FAPESP (proyeto tem\'atico 2001/07342-7).
\end{acknowledgements}

\begin{table*}
\begin{tabular}{r|rrrrrrr}
Name & $\overline{m}_I$& $(V-I)_{\rm 0,SBF}$ &$S/N$ & $BG$ & $\Delta_{\rm GC}$ &$(m-M)$\\\hline\hline
IM4\_2\_LSB1 & 28.28 $\pm$ 0.37& 0.932 & 5.0 & 0.36 &  0.01 & 31.00 $\pm$ 0.39 \\IM1\_2\_2\_LSB1 & 28.13 $\pm$ 0.37& 0.920 &  1.9 & 0.12 &  0.02 & 30.91 $\pm$ 0.39 \\
\hline
\end{tabular}
\caption{Summary of the SBF data for the two new dE candidates from Sect.~\ref{search} with detectable SBF signal. See Mieske et al.~(\cite{Mieske06b}, paper III) and Sect.~\ref{SBF} of this paper for details on the SBF measurement procedure. $BG$ gives the fraction of sky background fluctuation present in the uncorrected original fluctuation image. $\Delta_{\rm GC}$ gives the contribution of undetected GCs to the total flucutation signal in the original fluctuation image. The calibration relation adopted is the steep branch of case D in paper III, which is identical to the calibration by Tonry et al.~(\cite{Tonry01}). }
\label{sbfnew}
\end{table*}
\begin{longtable}{|l|rrrrrr|}
\caption{\label{sbfresults}Membership flags and photometric properties
  of investigated Fornax dE candidates, ordered by decreasing luminosity. Column 1
  gives the object identifier. FCC refers to the Fornax Cluster
  Catalog (FCC, Ferguson~\cite{Fergus89}). WF.. refers to new dE
  candidate detections from Hilker et al.~(\cite{Hilker03}, paper
  II) with the WFCCD camera. IM... refers to new dE candidate detections from this paper using IMACS. $V_{\rm 0}$ and $\mu_{\rm V,0}$ is from the IMACS data. The colour $(V-I)_{\rm 0}$ is the mean of the IMACS and WFCCD values, see text for more details. Errors are indicated in parentheses. The flags denote the various membership classifications as described in Sects.~\ref{SBF} to ~\ref{search}. Non-integer flags denote gradual classifications, see text. Galaxies marked with $^*$ have their photometry from paper II, because in the IMACS data the detection of those objects was unclear or they were resolved into several single sources. Note that all probable members (flag$<$ 2.5) do have IMACS photometry. New dE candidates marked by $^{**}$
  were not imaged in the WFCCD data from paper II, but are 
  listed in the FCC as likely members. The galaxy marked with $^{***}$
  is listed in the FCC as likely background / possible cluster member
  under FCC B1241 and in Hilker et al.~(\cite{Hilker99}) as
  GCF\_1-44. The four new dE candidates marked by $^+$ were detected
  only by visual inspection, not by SExtractor.}\vspace{-0.2cm}\\\hline\hline
Name & Flag & $V_{\rm 0}$ & $\mu_{\rm V,0}$ & $(V-I)_{\rm 0}$ & RA [2000] & DEC [2000]\\
\hline
\endfirsthead
\caption{continued.}\\
\hline\hline
Name & Flag & $V_{\rm 0}$ & $\mu_{\rm V,0}$ & $(V-I)_{\rm 0}$ & RA [2000] & DEC [2000]\\
\hline
\endhead
\hline
\endfoot
FCC~222                                  &     1. &     14.80 (0.01)   &     21.54 (0.08)   &     1.112 (0.040)   &   3:39:13.31 &   $-$35:22:16.43 \\ 
FCC~188                                  &     1. &     15.60 (0.01)   &     21.84 (0.05)   &     1.102 (0.024)   &   3:37:04.68 &   $-$35:35:24.36 \\ 
FCC~211                                  &     1. &    15.74 (0.01)   &     20.77 (0.06)   &     1.027 (0.028)   &   3:38:21.65 &   $-$35:15:34.99 \\ 
FCC~223                                  &     1. &    15.89 (0.02)   &     22.77 (0.07)   &     0.997 (0.032)   &   3:39:19.63 &   $-$35:43:27.98 \\ 
FCC~241                                  &     1. &    16.22 (0.02)   &     22.97 (0.08)   &     1.012 (0.040)   &   3:40:23.63 &   $-$35:16:35.15 \\ 
FCC~274                                  &     1. &    16.31 (0.01)   &     22.34 (0.08)   &     0.887 (0.040)   &   3:42:17.30 &   $-$35:32:26.84 \\ 
FCC~156                                  &     1. &    16.85 (0.01)   &     23.19 (0.06)   &     0.997 (0.028)   &   3:35:42.80 &   $-$35:20:18.13 \\ 
FCC~208                                  &     1. &    17.23 (0.02)   &     23.02 (0.14)   &     1.012 (0.069)   &   3:38:18.88 &   $-$35:31:50.81 \\ 
FCC~196                                  &     1. &     17.30 (0.02)   &     22.91 (0.08)   &     0.992 (0.040)   &   3:37:34.13 &   $-$35:49:44.44 \\ 
FCC~160                                  &     1. &    17.34 (0.01)   &     23.24 (0.09)   &     0.992 (0.044)   &   3:36:04.16 &   $-$35:23:19.18 \\ 
FCC~194                                  &     1. &     17.80 (0.02)   &     22.52 (0.09)   &     0.997 (0.044)   &   3:37:18.01 &   $-$35:41:57.08 \\ 
FCC~287                                  &     1. &    17.89 (0.02)   &     22.93 (0.16)   &     0.897 (0.077)   &   3:43:13.60 &   $-$35:31:07.21 \\ 
FCC~269                                  &     2. &    18.21 (0.09)   &     24.75 (0.17)   &     1.042 (0.081)   &   3:41:57.06 &   $-$35:17:39.05 \\ 
FCC~140                                  &     1. &    18.25 (0.02)   &     23.75 (0.09)   &     0.947 (0.044)   &   3:34:56.59 &   $-$35:11:26.99 \\ 
FCC~218                                  &    1.&    18.25 (0.05)   &      23.50 (0.17)   &     0.932 (0.081)   &   3:38:45.53 &   $-$35:15:58.86 \\ 
FCC~226                                  &     2. &    18.41 (0.29)   &     26.12 (0.45)   &     1.027 (0.218)   &   3:39:50.41 &   $-$35:01:21.14 \\ 
FCC~228                                  &     2. &    18.69 (0.08)   &     24.51 (0.54)   &     1.007 (0.263)   &   3:39:51.48 &   $-$35:19:20.68 \\ 
FCC~229                                  &     1. &    18.75 (0.04)   &     24.37 (0.12)   &     0.902 (0.057)   &   3:39:55.33 &   $-$35:39:42.30 \\ 
FCC~168                                  &     1. &    18.78 (0.05)   &     23.66 (0.16)   &     0.917 (0.077)   &   3:36:28.06 &   $-$35:12:38.66 \\ 
FCC~157                                  &     2. &    18.85 (0.04)   &     24.66 (0.10)   &     1.022 (0.048)   &   3:35:42.83 &   $-$35:30:50.69 \\ 
FCC~141                                  &     3. &    18.89 (0.01)   &     21.64 (0.04)   &     0.752 (0.020)   &   3:34:57.10 &   $-$35:12:18.00 \\ 
FCC~214                                  &     1. &    18.91 (0.04)   &      23.30 (0.20)   &     0.932 (0.097)   &   3:38:36.66 &   $-$35:50:03.16 \\ 
FCC~197                                  &    1.7 &    18.94 (0.04)   &     23.91 (0.18)   &     0.917 (0.085)   &   3:37:41.17 &   $-$35:17:45.13 \\ 
FCC~154                                  &    1.3 &    18.99 (0.03)   &     24.25 (0.07)   &     1.007 (0.032)   &   3:35:30.49 &   $-$35:15:06.44 \\ 
FCC~284                                  &    1.7 &    19.07 (0.08)   &     24.62 (0.39)   &     0.722 (0.190)   &   3:42:55.16 &   $-$35:20:37.25 \\ 
FCC~145                                  &     1. &    19.07 (0.02)   &     23.57 (0.15)   &     0.982 (0.073)   &   3:35:05.57 &   $-$35:13:05.81 \\ 
IM4\_2\_LSB1$^{**}$~(FCC~175)            &     1. &    19.12 (0.06)   &     24.14 (0.23)   &     0.942 (0.109)   &   3:36:43.03 &   $-$35:26:07.80 \\ 
FCC~215                                  &     1. &    19.21 (0.06)   &     24.46 (0.20)   &     0.822 (0.097)   &   3:38:37.57 &   $-$35:45:24.70 \\ 
FCC~191                                  &     1. &    19.26 (0.05)   &     23.42 (0.26)   &     0.967 (0.125)   &   3:37:10.07 &   $-$35:23:11.51 \\ 
FCC~144                                  &     1. &    19.26 (0.04)   &     24.05 (0.14)   &     0.937 (0.069)   &   3:35:00.29 &   $-$35:19:21.83 \\ 
FCC~220                                  &     2. &     19.30 (0.08)   &     24.14 (0.33)   &     1.137 (0.158)   &   3:38:55.32 &   $-$35:14:11.29 \\ 
IM7\_2\_LSB1$^{**}$~$^+$(FCC~272)        &     2. &    19.45 (0.11)   &     25.17 (0.28)   &     0.972 (0.137)   &   3:42:11.20 &   $-$35:26:34.44 \\ 
FCC~275                                  &     2. &    19.61 (0.09)   &     24.69 (0.15)   &     0.747 (0.073)   &   3:42:19.14 &   $-$35:33:39.60 \\ 
WFLSB4\_4                                &     2. &    19.63 (0.09)   &     25.08 (0.18)   &     0.882 (0.089)   &   3:36:59.84 &   $-$35:20:37.75 \\ 
IM1\_6\_LSB2$^{**}$~$^+$(FCC~127)        &     2. &    19.67 (0.04)   &     24.54 (0.07)   &     0.902 (0.036)   &   3:34:06.03 &   $-$35:16:37.92 \\ 
FCC~192                                  &     1. &     19.70 (0.08)   &     24.33 (0.39)   &     1.007 (0.190)   &   3:37:10.35 &   $-$35:53:17.34 \\ 
IM1\_2\_5\_LSB1$^{***}$~$^+$~(FCC~B1241) &     2. &    19.72 (0.10)   &       24.00 (0.28)   &     0.982 (0.137)   &   3:38:42.23 &   $-$35:33:08.28 \\ 
WFLSB6-4                                 &     1. &    19.75 (0.04)   &     23.53 (0.10)   &     0.972 (0.048)   &   3:35:57.93 &   $-$35:20:53.84 \\ 
IM1\_2\_2\_LSB1                          &    1.3 &     19.80 (0.08)   &     22.96 (0.23)   &     0.892 (0.113)   &   3:39:22.49 &   $-$35:35:24.83 \\ 
FCC~401                                  &     2. &    19.84 (0.14)   &     25.25 (0.32)   &     0.742 (0.154)   &   3:38:57.67 &   $-$35:11:03.88 \\ 
WFLSB9-4                                 &     1. &    19.92 (0.04)   &     23.49 (0.27)   &     0.857 (0.129)   &   3:38:37.23 &   $-$35:55:02.60 \\ 
WFLSB1-3                                 &     2. &    19.99 (0.06)   &     24.71 (0.13)   &     0.807 (0.065)   &   3:39:41.62 &   $-$35:31:53.26 \\ 
WFLSB11-4                                &     2. &    20.03 (0.16)   &     24.73 (0.25)   &     0.907 (0.121)   &   3:42:25.27 &   $-$35:35:41.17 \\ 
WFLSB10-4                                &    2.7 &    20.08 (0.09)   &     23.77 (0.26)   &     1.002 (0.125)   &   3:37:03.46 &   $-$35:48:02.16 \\ 
IM1\_1\_7\_LSB1                          &     2. &    20.16 (0.06)   &     23.41 (0.13)   &     0.792 (0.065)   &   3:38:04.08 &   $-$35:16:55.49 \\ 
IM1\_6\_LSB1$^{**}$~(FCC~131)            &     2. &    20.17 (0.05)   &     24.22 (0.17)   &     0.907 (0.081)   &   3:34:12.35 &   $-$35:13:41.99 \\ 
WFLSB11-3$^*$                            &     3. &    20.21 (0.25)   &     25.96 (0.06)   &     0.772 (0.070)   &   3:42:33.07 &   $-$35:31:56.60 \\ 
WFLSB6-2                                 &     1. &    20.21 (0.04)   &      24.30 (0.20)   &     0.842 (0.097)   &   3:34:57.74 &   $-$35:13:23.81 \\ 
WFLSB6-1                                 &    3.3 &    20.23 (0.04)   &     23.51 (0.11)   &     0.772 (0.053)   &   3:35:06.06 &   $-$35:06:25.16 \\ 
IM7\_6\_LSB1                             &    2.3 &    20.25 (0.07)   &      23.40 (0.16)   &     0.602 (0.077)   &   3:41:34.22 &   $-$35:24:09.90 \\ 
WFLSB10-7                                &     2. &    20.26 (0.18)   &     25.28 (0.30)   &     1.062 (0.145)   &   3:37:41.98 &   $-$35:57:18.36 \\ 
WFLSB6-6$^*$                             &     4. &    20.34 (0.15)   &     25.86 (0.04)   &     1.612 (0.050)   &   3:35:49.64 &   $-$35:07:21.86 \\ 
WFLSB2-1                                 &     2. &     20.40 (0.16)   &     25.14 (0.36)   &     0.817 (0.174)   &   3:37:38.71 &   $-$35:23:07.30 \\ 
WFLSB10-6                                &    1.7 &     20.40 (0.13)   &     24.73 (0.40)   &     0.997 (0.194)   &   3:37:27.50 &   $-$35:57:46.44 \\ 
WFLSB10-8                                &     1. &    20.42 (0.10)   &     24.68 (0.22)   &     0.902 (0.105)   &   3:38:06.26 &   $-$36:03:01.08 \\ 
WFLSB1-6                                 &     2. &    20.45 (0.08)   &     25.24 (0.21)   &     0.777 (0.101)   &   3:39:56.36 &   $-$35:37:19.85 \\ 
WFLSB11-2                                &     2. &    20.54 (0.10)   &     24.22 (0.27)   &     0.797 (0.129)   &   3:42:43.02 &   $-$35:32:26.16 \\ 
WFLSB5-1                                 &     2. &    20.54 (0.11)   &     25.53 (0.30)   &     0.922 (0.145)   &   3:35:05.38 &   $-$35:27:03.38 \\ 
WFLSB4-3$^*$                             &     4. &     20.60 (0.22)   &     26.19 (0.07)   &     1.082 (0.090)   &   3:37:05.99 &   $-$35:20:31.02 \\ 
WFLSB1-5                                 &     2. &    20.62 (0.16)   &     26.45 (0.24)   &     1.007 (0.117)   &   3:40:02.76 &   $-$35:27:54.50 \\ 
WFLSB1-2                                 &     2. &    20.65 (0.69)   &     26.86 (0.78)   &     0.792 (0.380)   &   3:39:28.77 &   $-$35:34:21.90 \\ 
WFLSB2-6                                 &     3. &    20.71 (0.09)   &     23.45 (0.23)   &     1.632 (0.109)   &   3:37:04.65 &   $-$35:40:48.76 \\ 
IM1\_2\_6\_LSB1$^+$                      &     2. &    20.78 (0.14)   &     23.66 (0.23)   &     0.722 (0.109)   &   3:38:47.54 &   $-$35:43:32.52 \\ 
WFLSB11-8$^*$                            &     4. &    20.86 (0.33)   &     26.93 (0.10)   &     1.442 (0.120)   &   3:42:52.87 &   $-$35:33:16.52 \\ 
WFLSB2-7                                 &    2.3 &    20.94 (0.11)   &     24.97 (0.18)   &     0.917 (0.085)   &   3:37:44.44 &   $-$35:42:12.92 \\ 
WFLSB3-4$^*$                             &     4. &    20.98 (0.20)   &     26.46 (0.11)   &     0.632 (0.130)   &   3:40:18.98 &   $-$35:08:16.76 \\ 
WFLSB3-2                                 &    1.7 &      21.00 (0.12)   &     24.45 (0.26)   &     0.982 (0.125)   &   3:39:07.18 &   $-$35:07:22.33 \\ 
WFLSB10-2                                &     2. &    21.02 (0.14)   &     24.99 (0.27)   &     0.932 (0.129)   &   3:36:54.31 &   $-$35:44:46.32 \\ 
WFLSB1-1                                 &    2.7 &    21.02 (0.07)   &     23.31 (0.11)   &     0.722 (0.053)   &   3:39:58.52 &   $-$35:33:23.80 \\ 
WFLSB6-3                                 &     3. &    21.07 (0.05)   &     23.33 (0.09)   &     0.632 (0.044)   &   3:34:58.37 &   $-$35:15:29.74 \\ 
WFLSB11-1                                &    2.3 &    21.15 (0.14)   &     24.13 (0.26)   &     0.717 (0.125)   &   3:42:17.84 &   $-$35:28:18.84 \\ 
WFLSB5-6                                 &    1.3 &    21.26 (0.08)   &     24.41 (0.18)   &     0.932 (0.085)   &   3:35:47.44 &   $-$35:21:41.64 \\ 
WFLSB3-5                                 &     2. &    21.33 (0.18)   &     25.27 (0.31)   &     0.747 (0.150)   &   3:39:29.55 &   $-$35:04:46.97 \\ 
WFLSB2-3                                 &     2. &    21.45 (0.26)   &     25.61 (0.25)   &     1.007 (0.121)   &   3:37:15.72 &   $-$35:21:26.50 \\ 
WFLSB12-2                                &     2. &    21.45 (0.30)   &     26.55 (0.33)   &     0.492 (0.158)   &   3:43:02.06 &   $-$35:19:39.72 \\ 
WFLSB2-13$^*$                            &     4. &    21.48 (0.18)   &     25.81 (0.03)   &     0.862 (0.040)   &   3:37:51.20 &   $-$35:21:45.36 \\ 
WFLSB10-1                                &     2. &    21.52 (0.22)   &     25.04 (0.32)   &     0.807 (0.154)   &   3:38:08.40 &   $-$36:00:30.60   \\ 
WFLSB9-5                                 &    2.7 &    21.53 (0.18)   &     24.14 (0.43)   &     1.872 (0.210)   &   3:38:35.47 &   $-$35:55:06.49 \\ 
WFLSB1-9$^*$                             &     4. &    21.55 (0.25)   &     25.76 (0.07)   &     0.452 (0.080)   &   3:38:41.70 &   $-$35:23:29.52 \\ 
WFLSB6-8                                 &     2. &    21.62 (0.18)   &     24.98 (0.38)   &     0.977 (0.186)   &   3:36:01.80 &   $-$35:18:37.99 \\ 
WFLSB1-7$^*$                             &     3. &    21.62 (0.55)   &     25.96 (0.09)   &     0.752 (0.110)   &   3:39:42.62 &   $-$35:44:42.83 \\ 
WFLSB4-6$^*$                             &     3. &    21.67 (0.36)   &     26.96 (0.14)   &     0.982 (0.170)   &   3:38:02.25 &   $-$35:16:46.99 \\ 
WFLSB7-3$^*$                             &     4. &    21.89 (0.39)   &     25.96 (0.07)   &     1.082 (0.080)   &   3:40:45.30 &   $-$35:30:07.02 \\ 
WFLSB2-11$^*$                            &     3. &    21.89 (0.66)   &     26.93 (0.38)   &    -0.018 (0.450)   &   3:36:40.68 &   $-$35:35:21.84 \\ 
WFLSB12-3                                &     2. &    21.92 (0.37)   &     26.23 (0.58)   &     0.557 (0.283)   &   3:43:01.63 &   $-$35:15:01.80 \\ 
WFLSB2-10$^*$                            &    2.7 &    21.94 (0.28)   &     25.56 (0.11)   &     0.307 (0.053)   &   3:37:49.95 &   $-$35:32:58.67 \\ 
WFLSB1-4$^*$                             &     4. &    21.96 (0.40)   &     25.89 (0.07)   &     0.612 (0.090)   &   3:39:41.97 &   $-$35:28:04.98 \\ 
WFLSB2-9                                 &     2. &    21.98 (0.48)   &     26.11 (0.78)   &     1.067 (0.376)   &   3:38:06.89 &   $-$35:35:30.01 \\ 
WFLSB9-6                                 &    2.7 &    22.04 (0.30)   &     25.29 (0.39)   &     0.812 (0.190)   &   3:38:54.31 &   $-$35:59:11.36 \\ 
WFLSB11-7                                &     2. &    22.05 (0.26)   &     25.85 (0.38)   &     0.867 (0.182)   &   3:42:47.38 &   $-$35:23:21.34 \\ 
WFLSB10-5                                &     3. &    22.25 (0.23)   &     24.17 (0.21)   &     1.017 (0.101)   &   3:36:57.19 &   $-$35:50:11.40 \\ 
IM3\_7\_LSB1                             &     2. &    22.31 (0.39)   &     25.29 (0.64)   &     1.132 (0.311)   &   3:36:45.93 &   $-$35:55:27.19 \\ 
IM3\_7\_LSB2                             &     2. &    22.31 (0.30)   &     25.56 (0.37)   &     1.022 (0.178)   &   3:36:54.99 &   $-$35:56:39.84 \\ 
WFLSB6-7                                 &    2.3 &    22.35 (0.15)   &      25.80 (0.28)   &     0.747 (0.137)   &   3:35:38.25 &   $-$35:05:41.20 \\ 
WFLSB10-9                                &     2. &    22.35 (0.34)   &     25.86 (0.60)   &     1.187 (0.291)   &   3:38:10.70 &   $-$35:53:54.24 \\ 
WFLSB7-4$^*$                             &     3. &    22.38 (0.29)   &     25.41 (0.18)   &     0.002 (0.210)   &   3:41:46.77 &   $-$35:30:56.95 \\ 
WFLSB2-12                                &    2.7 &    22.41 (0.37)   &     25.67 (0.38)   &     1.222 (0.186)   &   3:37:30.07 &   $-$35:22:56.17 \\ 
WFLSB1-11$^*$                            &     4. &    22.42 (0.55)   &     25.73 (0.07)   &     1.012 (0.080)   &   3:39:19.84 &   $-$35:32:13.27 \\ 
IM3\_3\_LSB1                             &     2. &     22.60 (0.37)   &     25.36 (0.40)   &     1.542 (0.194)   &   3:37:47.22 &   $-$35:44:44.23 \\ 
WFLSB2-4                                 &    2.7 &    22.78 (0.38)   &     24.76 (0.39)   &     1.277 (0.190)   &   3:37:09.33 &   $-$35:26:46.82\\\hline
\end{longtable}
\end{document}